\documentclass[11pt]{article} 
\pdfoutput=1  
\usepackage[top=1in, bottom=1in, left=1in, right=1in]{geometry}
\usepackage{amsmath}
\usepackage{amssymb}
\usepackage{graphicx}
\usepackage[export]{adjustbox}
\usepackage[format=plain,justification=centering,singlelinecheck=false,font=small,labelfont=bf,labelsep=space]{caption}
\usepackage{bm}
\usepackage{subfig}
\usepackage[usenames,dvipsnames]{color}
\usepackage[hyperfootnotes=false]{hyperref}
\usepackage{cleveref}

\usepackage[utf8]{inputenc}
\usepackage[backend=biber,style=nature]{biblatex}
\usepackage{setspace}

\usepackage{xspace}
\usepackage{enumitem}
\usepackage{wrapfig}
\usepackage{authblk}

\usepackage[outdir=./]{epstopdf}
\usepackage{longtable}
\usepackage{multicol}
\usepackage{ragged2e}
\DeclareRefcontext{supp}{labelprefix=S} 

\title{Cross-scale covariance for material property prediction}

\author[1]{Benjamin A. Jasperson\footnote{corresponding author}}
\author[2]{Ilia Nikiforov}
\author[3]{Amit Samanta}
\author[3]{Fei Zhou}
\author[2]{Ellad B. Tadmor}
\author[3]{Vincenzo Lordi}
\author[3]{Vasily V. Bulatov}

\affil[1]{University of Illinois Urbana-Champaign}
\affil[2]{Department of Aerospace Engineering and Mechanics, University of Minnesota, Minneapolis, MN 55455, USA}
\affil[3]{Lawrence Livermore National Laboratory}

\date{\vspace{-1cm}}

\begin{document}

\maketitle

\begin{abstract}
	A simulation can stand its ground against experiment only if its prediction uncertainty is known. The unknown accuracy of interatomic potentials (IPs) is a major source of prediction uncertainty, severely limiting the use of large-scale classical atomistic simulations in a wide range of scientific and engineering applications. Here we explore covariance between predictions of metal plasticity, from 178 large-scale ($\sim 10^8$ atoms) molecular dynamics (MD) simulations, and a variety of indicator properties computed at small-scales ($\leq 10^2$ atoms). All simulations use the same 178 IPs. In a manner similar to statistical studies in public health, we analyze correlations of strength with indicators, identify the best predictor properties, and build a cross-scale ``strength-on-predictors'' regression model. This model is then used to quantify uncertainty over the statistical pool of IPs. Small-scale predictors found to be highly covariant with strength are computed using expensive quantum-accurate calculations and used to predict flow strength, within the uncertainty bounds established in our statistical study.
	
\end{abstract}

\begin{refsection}

With steady growth and decreasing cost of high-performance computing (HPC), atomistic simulations are becoming an increasingly practical component of materials R\&D. Classical atomistic simulations rely on computationally expedient interatomic potentials (IPs) that permit material models with many millions, billions or even trillions of atoms.  The availability of accurate, transferable and computationally efficient IPs is a key ingredient for continued successful applications of atomistic simulations in materials research.  Ongoing explosive developments in machine-learned IPs promise to bring the accuracy of atomistic simulations closer in line with that of ground-truth (GT) quantum mechanical calculations. Given their high cost and unfavorable scaling, GT calculations are limited to just a few hundred atoms, thereby defining the scales at which GT reference data is presently computed.  What remains unclear is whether training an IP on small-scale GT data, however exhaustive and accurate, can translate into similarly accurate atomistic predictions at scales far beyond the reach of GT calculations, where simulations with IPs are the only option.

As an alternative to the presently impossible task of validating large-scale IP-based atomistic simulations against GT, here we propose to assess the prediction accuracy by quantifying its covariance with separate small-scale ``indicator'' properties calculated using the same IP, and identifying the best ``predictors'' among them. These predictors are at a scale where direct validation against GT is indeed possible and available. The logic of our proposed approach is analogous to uses of statistical inference for forecasting in financial engineering, public health, operations research, etc. For example, statistical studies in public health are used to predict a quantifiable health outcome (e.g., blood sugar level at 60 years of age or patient longevity) from personal traits (indicators) of the same patient (e.g., body mass index, gender, ethnicity, and so on). Despite the inherent uncertainty in predicting future health outcomes for any specific individual, results of such studies performed over a statistically representative pool of patients reveal correlations and trends that inform the general population and medical professionals of quantifiable health risks. Extending this to atomistic simulations, to the extent that statistical covariance between a large-scale quantity of interest (QoI) (health outcome) and small-scale indicator properties (personal traits) can be established, a regression model can be constructed and used to predict the QoI from the most covariant indicators (predictors) computed at small scales where GT calculations are feasible. Furthermore, the same statistics can be used to estimate the uncertainty of such predictions.

\begin{figure}
    \centering
    \includegraphics[width=14cm]{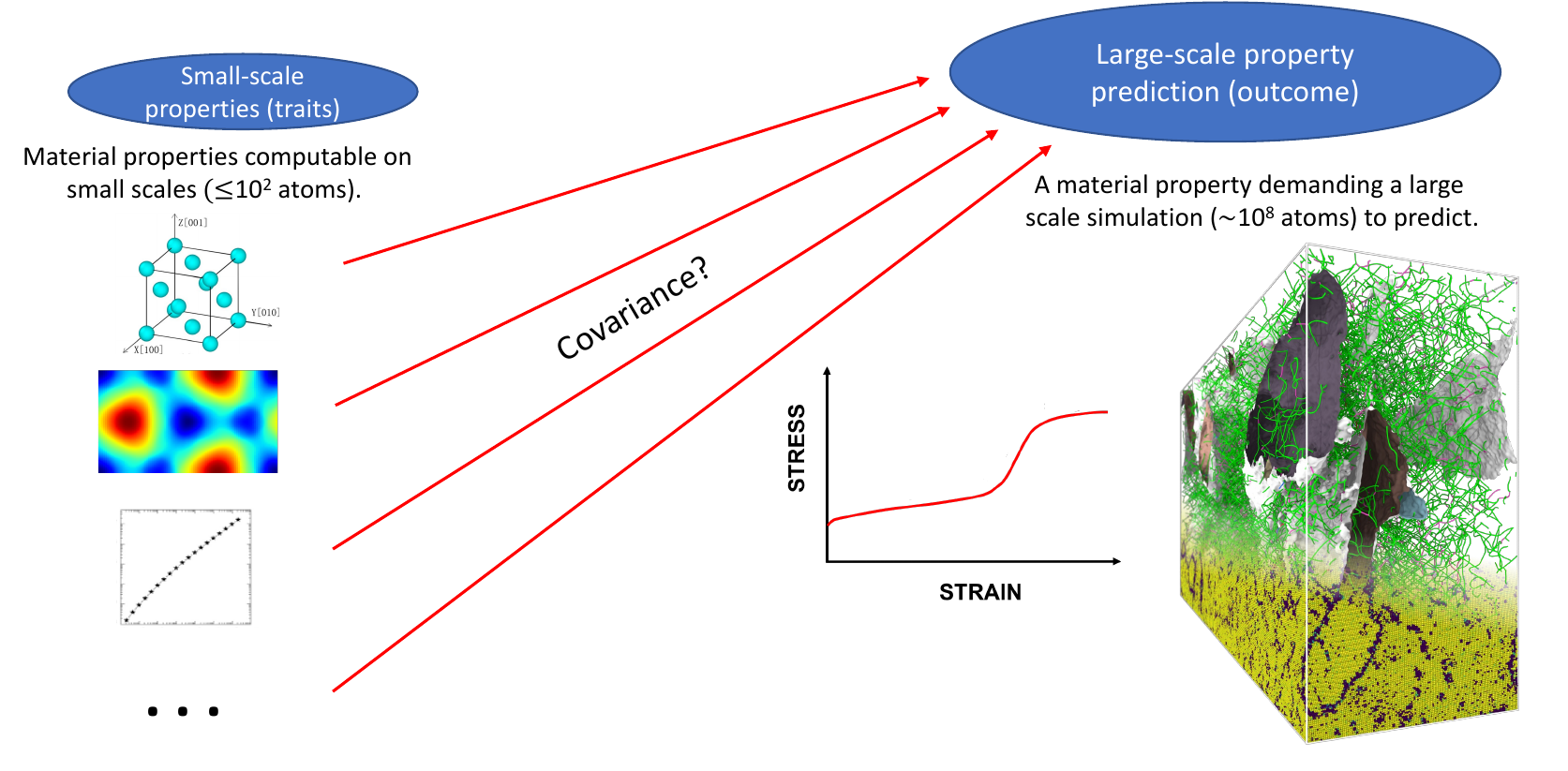}
    \caption{Cross-scale covariance is defined as the covariance between a material QoI requiring a large-scale simulation to compute, e.g., plastic flow strength, and other properties of the same material that can be computed at much smaller scales where GT predictions are presently feasible.
    }
    \label{fig:overview}
\end{figure}

As a test bed in the material simulation context, we examine the covariance of plastic flow strength in metals (our QoI) computed in large-scale molecular dynamics (MD) simulations with indicator properties of the same materials computable on small scales, e.g., elastic constants, surface energies, etc.\ (\Cref{fig:overview}). We perform the study over a pool of 178 IPs (statistical samples) for face-centered cubic (FCC) metals. The IPs used are available in the OpenKIM repository at \url{https://openkim.org} \cite{TES11} along with their predictions for a range of material properties computed using robust and vetted computational protocols. As described in the Methods section, large-scale strength prediction involving tens of millions of atoms, and calculations of small-scale properties, were performed under identical conditions for all 178 IP models included in the statistical pool. Indicator properties found covariant with the computed MD strength become candidate independent predictor variables for building a ``strength-on-predictors'' regression function subsequently used to predict strength from GT theory along with uncertainty bounds obtained through the application of existing methods of statistical cross-validation.

We note that applicability of a regression model and associated uncertainty to predicting plastic flow strength from GT data is predicated on our starting assumption that the GT material model belongs to the same statistical pool as the IP models. Thus we assume that a (never-to-be-known) GT prediction of the QoI is covariant with GT predictors in the same manner that those quantities are related for the IPs included in our study.
Our reasoning is that atomic trajectories defining a large-scale material QoI, such as the plastic flow strength, are sampling some of the same regions of a model's potential energy surface (PES) (i.e., potential energy as a function of the atomic configuration) that are also sampled in computing small-scale (predictor) properties of the material model. Of course, only a small sampling of the PES is accounted for in computing any one of the predictor properties but, taken together, such small samples reflect the overall PES structure in the regions of interest for a given QoI. Thus our expectation that DFT ``belongs to the same statistical pool'' as the IPs, is a statement that (1) the QoI computed via DFT is tied to the same PES regions as the QoI computed via IPs; and (2) that DFT shares a similar PES structure to the IPs in these regions and therefore correlates similarly with the same predictor properties.

\section*{Simulation results and data for statistical analyses}
\label{sec:results}

Our plan was to run MD simulations of crystal strength with as many IPs as we could possibly afford, compute the covariance of the dependent variable (predicted strength values) with a set of independent variables (small-scale indicator properties) extracted from the OpenKIM repository for the same IPs, and build a statistical regression model relating strength to indicators found to be most covariant. However, collecting the data presented challenges. MD simulations using one particular IP could not be completed due to numerical instabilities encountered irrespective of the integration time step. Another 14 IPs were excluded after subsequent analyses revealed anomalies or irregularities in crystal response to straining, such as crystal rotation, phase transformations and formation of amorphous phases, or voids.  All such unexpected behaviors were accompanied by a strength response deviating qualitatively from that of the remaining 163 IP models. Although decisions to exclude 15 out of 178 IPs relied on expert judgment, the reduction was unbiased, as the irregular results were taken out of the pool before any statistical analyses were undertaken.

At the same time, inspection of small-scale indicator properties of the remaining 163 IPs, precomputed in the OpenKIM repository, revealed some missing or out-of-bounds values deemed unreliable.
Out of 62 initially considered indicators, 27 properties contained numerous missing or unreliable values and were excluded from our study altogether.
To fill the gaps in the $163 \times 35$ table of the remaining small-scale properties, wherever possible we 1) computed the missing property values, and 2) re-computed property values falling outside of their physically reasonable range. Still, some properties for several IPs could not be re-computed or remained unreasonable: all such IP samples were excluded from computing covariance of strength with their missing small-scale properties (see Supplementary Information (SI) for details).

\begin{figure}[th!]
    \centering
    \subfloat[\centering]
        {\includegraphics[width=3in]{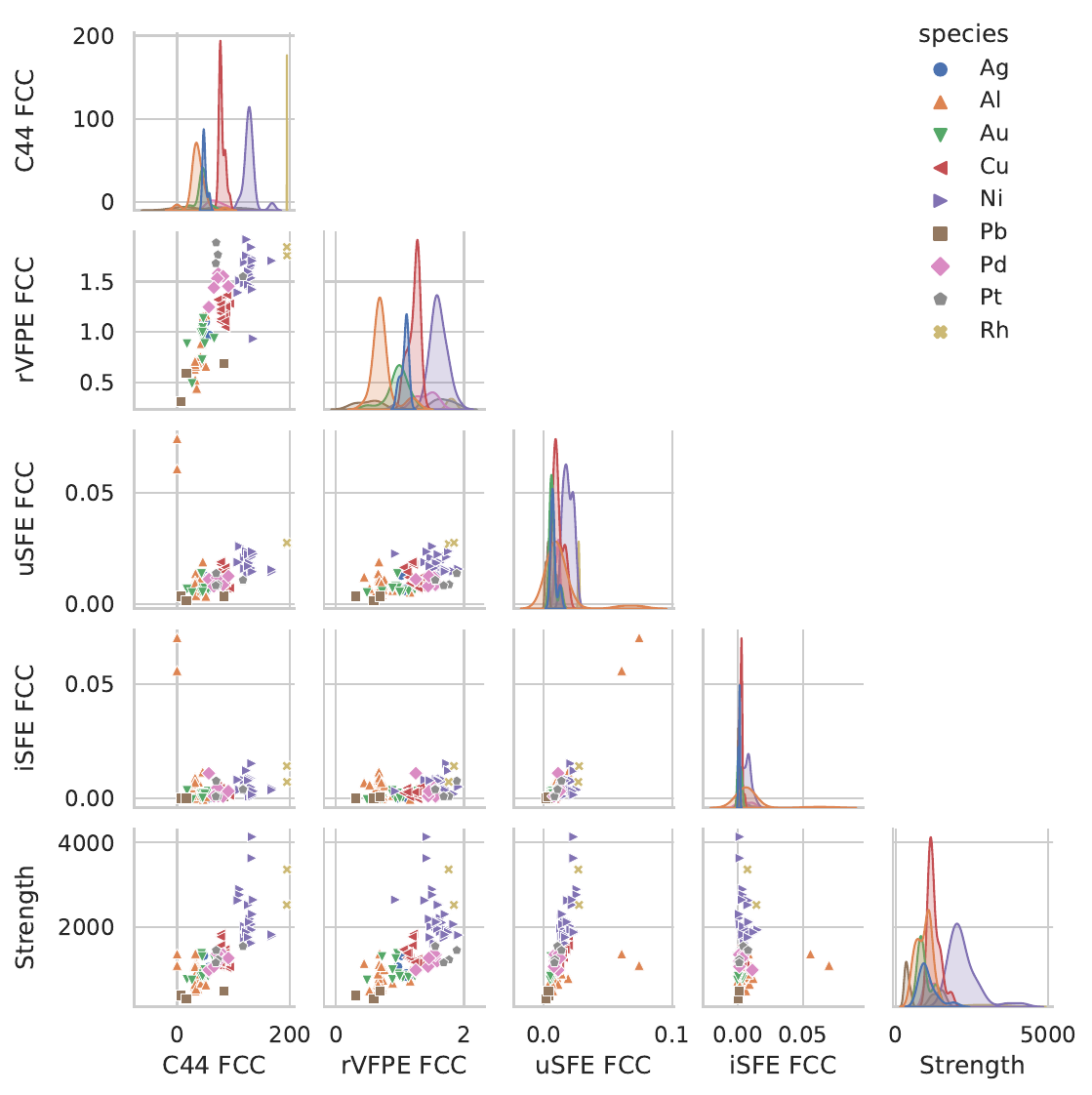}
         \label{fig:pairplot}}
    \subfloat[\centering]
        {\includegraphics[width=2in]{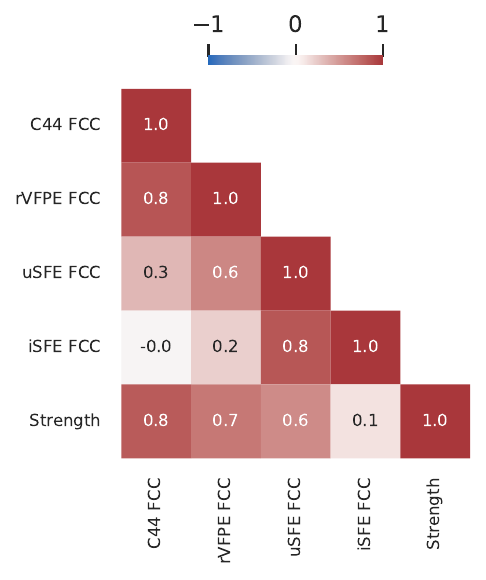}
         \label{fig:corr_plot}}\\
    \subfloat[\centering]
        {\includegraphics[height=2.5in]{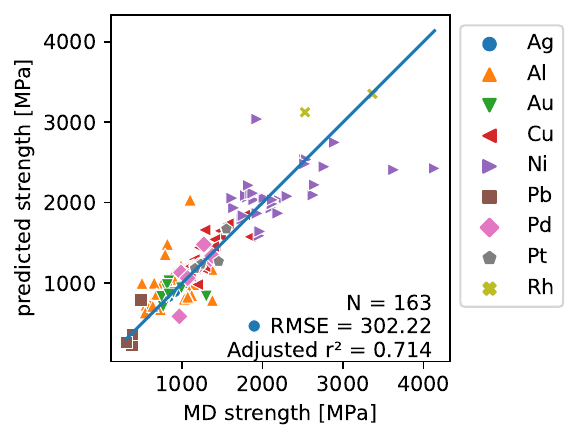}
         \label{fig:linear_all_pred}}
    \subfloat[\centering]
        {\includegraphics[height=2.5in]{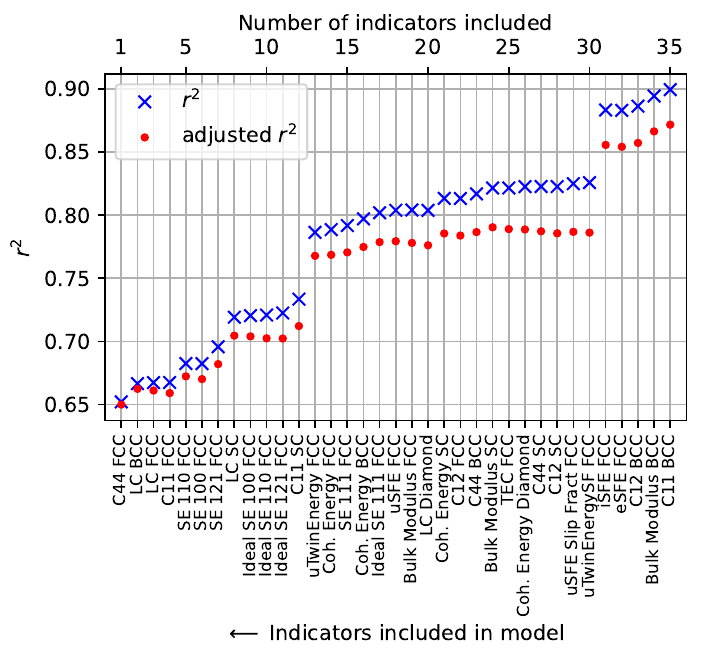}
         \label{fig:linear_all_r2}}
    \caption{(a) Scatter plots of correlations of small-scale indicator properties and strength evaluated on the pool of 163 IPs for nine FCC metals. C44 = shear modulus (GPa), rVFE = Relaxed Vacancy Formation Energy (eV), uSFE = unstable Stacking Fault Energy (ev/A\textsuperscript{2}), iSFE = intrinsic Stacking Fault Energy (ev/A\textsuperscript{2}). (b) Heatmap of correlation coefficients showing highly correlated ($C_{44}$), moderately correlated (rVFE and uSFE), and minimally correlated (iSFE) properties. Highly correlated properties approach values of 1 or -1. (c) Predicted strength (using multi-linear regression and leave-one-out cross-validation, see text for details) versus  strength computed in MD simulations for the 163 IP models.
    (d) $r^2$ and adjusted $r^2$ goodness of fit as each of the 35 predictor properties are added in the order shown. Each point represents $r^2$ and adjusted $r^2$ of a model including the corresponding predictor and all predictors to its left. Here, the predictors are added in decreasing order of individual correlation with strength.}
    \label{fig:linear}
\end{figure}

In principle, any small-scale indicator property (an elastic constant, a surface energy, etc.) available in OpenKIM can be included in a regression model.
However to help select  \emph{predictors} for regression below, we first quantify covariance of each individual candidate property with the dependent variable, the computed MD strength.
The lower row in \Cref{fig:pairplot} presents scatter plots illustrating covariance of strength with four small-scale indicator properties computed with the same IPs, while the rows above show cross-covariance among the small-scale properties themselves (a more complete set of covariance plots is given in the SI).
To quantify pairwise covariance we use the standard Pearson’s correlation coefficient \cite{johnsonStatisticalReasoningMethods2004}: variables with correlation coefficient approaching 1 (positive correlation) or -1 (negative correlation) are highly correlated.
\Cref{fig:corr_plot} is a heatmap showing correlations of strength with and among the same small-scale indicator properties in \Cref{fig:pairplot}. (A more complete heatmap of correlations among all variables is given in the SI.)
Strength co-varies tightly with some indicators ($C_{44}$ elastic constant, unstable stacking fault energy (uSFE)), while showing little or no covariance with others.

The previous scatter plots and pairwise covariance calculations simply excluded the missing or out-of-bounds indicator values from the IP samples.
These missing values must be addressed when fitting a ``strength-on-predictors'' multi-variate regression model, as many procedures require a complete dataset.
One approach is to simply remove any IP samples with missing values; this would result in an unacceptable reduction of our statistical pool.
Here we use data imputation to replace missing and unreasonable indicator values with inferred values, based on a $k$-nearest neighbors approach (see SI for details). We again relied on expert opinion to bound ranges of reasonable values for small-scale property data; however, all decisions as to which property values to exclude or replace were made prior to statistical analyses.

We use multi-linear regression to illustrate construction of a statistical model relating flow strength to the indicator properties, leaving more advanced regression analyses to the SI. Prior to computing regression coefficients each variable in the regression model is shifted and scaled to have zero mean and unit variance \cite{pedregosaScikitlearnMachineLearning2011}.
\Cref{fig:linear_all_pred} presents a linear regression model for the flow strength built on 35 small-scale indicators properties extracted from the OpenKIM repository with scaling applied prior to imputing missing values for the 163 IPs included in the statistical pool.
\Cref{fig:linear_all_r2} is the plot of the standard goodness of fit parameter $r^2$ as a function of the number of predictors included in the multi-linear regression, in descending order of their correlation with strength.
Although $r^2$ rises monotonically with the number of included predictors (as it should), except for several notable increases, adding more predictors brings little improvement in the goodness of fit.
Shown in the same figure is the adjusted $r^2$ \cite{adjusted_r2} that normalizes goodness of fit for the number of independent variables and can decrease if the next included variable does not improve the fit significantly.
Plateaus in $r^2$ are observed when a group of highly collinear predictors are added to the expanding regression model.
Typically, such collinear groups include closely-related properties, e.g., elastic constants $C_{11}$, $C_{12}$ and $C_{44}$, that tend to have similar correlation coefficients with strength, resulting in them being added to the regression model in sequence.
The jumps occur whenever a predictor is added that is not highly collinear with the previously added predictors.
We note that although the adjusted $r^2$ reaches a high value of 0.87, a few outliers are still observed in the regression plot in \Cref{fig:linear_all_pred}.

\section*{Further data reduction and augmentation}

\begin{figure}[t]
    \centering
    \includegraphics[height=10cm]{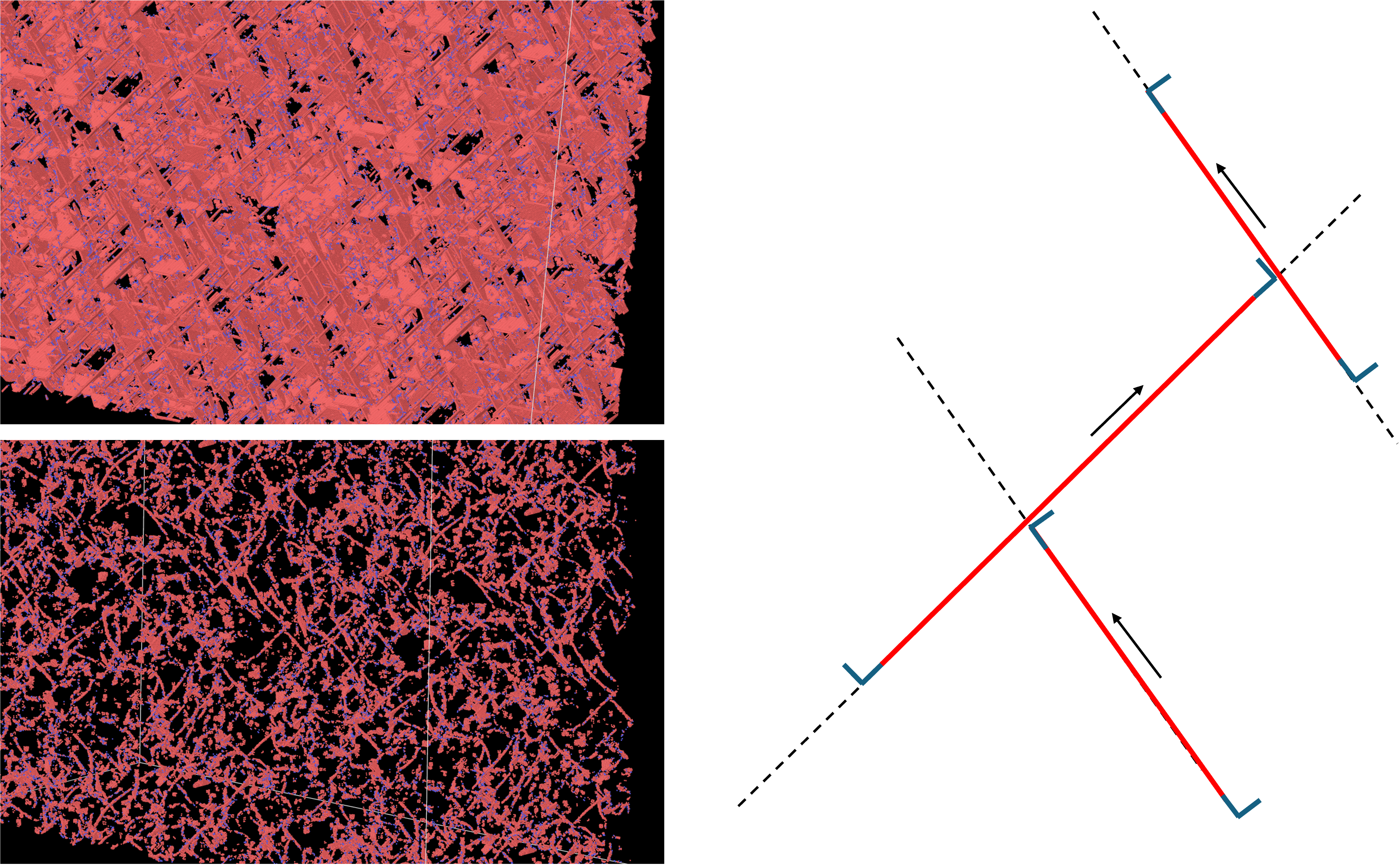}
    \caption{Thin slices through two plastically deformed crystals containing dislocations.  All atoms deemed ``perfect'' FCC by a classifier algorithm in Ovito \cite{stukowski2009visualization} are deleted for clarity leaving behind only a minority of atoms colored red deemed ``defective'' (classified as not FCC).  In an SF-jammed crystal, shown on the upper left, dislocations appear as ribbons of red atoms revealing widely dissociated (extended) dislocation cores delimiting an SF.  In the crystal on lower left, dislocations are only narrowly dissociated and appear as strings of defective atoms.  The 2D schematic on the right depicts SF jamming where three extended dislocations are shown on two symmetry-related glide planes (dashed lines).  Each dislocation consists of two Shockley partial dislocations (blue) connected by a ribbon of SF (red).  Arrows show directions of forces seen by each dislocation under externally applied mechanical stress. Dislocations can move along their respective glide planes, however their motion is blocked by wide SFs of other dislocations dissociated on intersecting glide planes. Jamming develops gradually as dislocations multiply under stress to a point where spacing between them becomes comparable to the width of their SFs (ribbons). In our set of 178 IP models, SF jamming largely affects models with low SF energy (which results in wide SFs).}
    \label{fig:jamming}
\end{figure}

Noting that some MD simulations predicted flow stress much greater than that predicted by the regression model in \Cref{fig:linear_all_pred} prompted us to additionally re-examine irregularities in all 178 IP models for which strength was computed in the MD. In doing so, an unusual mechanism was observed in some IP models, in which extended stacking faults block dislocation propagation on intersecting glide planes. Continued dislocation motion requires breaking through the SF obstacles \cite{yamakov2002dislocation}, thus necessitating higher stress than would be required if no such blocking obstacles were present. Termed here ``stacking fault jamming'' (SF jamming), this additional mechanism adds resistance to dislocation motion, thus contributing to an anomalously high plastic strength (\Cref{fig:jamming}). We performed additional visual attribution of all 178 models to two distinct types, ``SF jammed'' and ``not SF jammed.'' Of the 16 models which were identified as SF jammed, six had been previously excluded from the statistical pool based on other considerations. Outliers or not, we excluded all SF jammed IPs from further statistical analyses on the grounds that they exhibit a plastic response qualitatively different from the rest of the remaining IP models, thus paring our statistical pool down to 153 IP models.

Analysis of the defect microstructures observed in MD simulations suggested additional indicator properties not previously included as possible predictors. Even in the absence of SF jamming, crystal defects other than dislocations are observed in large quantities for all 153 IPs remaining in the pool. On further inspection, most non-dislocation defects were found to be vacancies and vacancy clusters likely produced by moving dislocations \cite{marian2004dynamic}. Leaving it to future work to establish specific atomistic mechanisms, the production of vacancies should add resistance to dislocation motion, thus increasing plastic strength. Following this logic, we developed new OpenKIM property calculations (KIM Test Drivers) for computing the vacancy formation energy (both relaxed and unrelaxed), the vacancy migration barrier, and the vacancy formation volume \cite{OpenKIM-TD:554849987965:001, OpenKIM-TD:647413317626:001}. Consistent with our observations, all four variables related to vacancies are found to have significant covariance with the flow strength. Thus, they were added as candidate independent variables in subsequent statistical analyses, bringing the number of available indicator properties to 39.

\section*{Down-selection of independent variables for regression}

Any indicator property computable on small scales can be considered a candidate predictor for cross-scale regression, such as material properties in GT repositories like Materials Project \cite{jainCommentaryMaterialsProject2013} and AFLOW \cite{curtarolo:setyawan:2012}, as well as energies, forces, stress, and other properties included in the rapidly expanding set of published datasets used in fitting IPs, such as those distributed through the ColabFit Exchange \cite{vita:fuemmeler:2023}. Ideally a cross-scale regression model should be built on indicators most covariant with the QoI. However, many indicator properties are highly collinear with each other. Adding an independent variable that is collinear with other independent variables that are already included in the regression model brings little additional information and makes the covariance matrix ill conditioned. For example, even though plastic flow strength is clearly covariant with the computed \{111\}, \{110\} and \{100\} surface energies, the latter energies are all collinear. Observing that most mutually collinear predictors belong to groups of related indicators (e.g., the group of three elastic constants), the number of independent regression variables can be reduced by including only one representative predictor property from each collinear property group. Furthermore, given the limited pool of available IP samples — now only 153 — we would like to preclude over-fitting by limiting the number of independent variables to a reasonable few. This is also preferable in reducing the effort required to compute predictors using computationally expensive GT methods such as density functional theory (DFT).

\begin{figure}[t!]
    \centering
    \subfloat[\centering]
        {\label{fig:factor_usage}
        \includegraphics{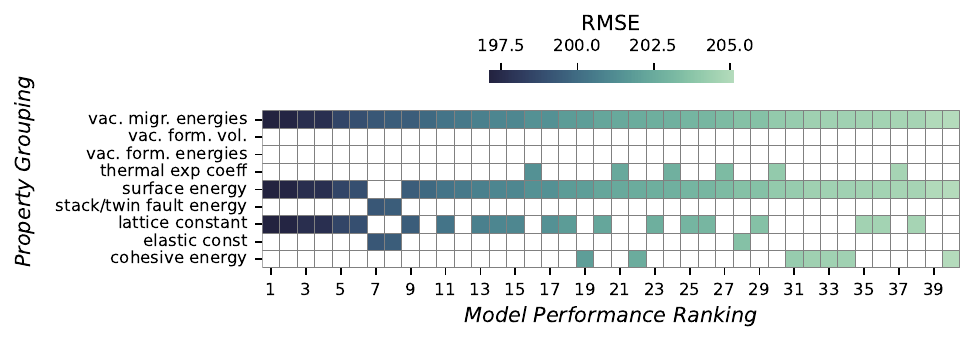}}
    \qquad
    \subfloat[\centering]
        {\label{fig:linear_3factors}
        \includegraphics{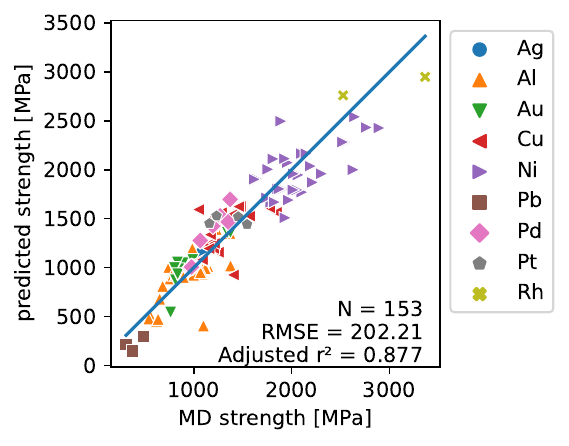}}
    \caption{(a) Heat map showing which generic predictor groups (e.g., ``elastic const''\ includes $C_{11}$, $C_{12}$ and $C_{44}$ for FCC, BCC, and SC) are used in the top performing regression models, as determined by repeated $k$-fold cross validation.  A filled box means at least one predictor from that group was used in the model. The color map at the top represents the root mean square error (RMSE) in the predicted strength in MPa. See the SI for factors included in each group. (b) Comparison of the strength predicted by a top multi-linear regression model with the  strength obtained from a large-scale MD simulation. The regression model is based on three predictors: \{111\} surface energy, lattice constant, and vacancy migration energy, all for FCC. The RMSE of the regression model is estimated using leave-one-out cross-validation, see the text for details.}
\end{figure}

 To select the most informative indicators, we use repeated $k$-fold cross-validation \cite{celisse2014optimal} to evaluate all 9919 regression models built on one, two and three out of 39 candidate indicators (39 choose 1 + 39 choose 2 + 39 choose 3). The $k$-fold method is useful in evaluating the ability of a model to predict previously unseen data especially when data itself is in short supply. Here we randomly divide our statistical pool of 153 IP samples into 10 groups or folds of (nearly) equal size and reserve each fold once for testing while using the remaining nine folds for training. For every combination of indicators, random partitioning into 10 folds is repeated 5 times. In \Cref{fig:factor_usage}, 3-variable regression models are shown in decreasing order of their resulting goodness of fit $r^2$ and predictors are lumped together into similar groups (see the SI for the predictors included in each group).
 The frequency of appearance of a given predictor group in the top performing regression models is an indication of the group’s importance in predicting strength. Consistent with its highest correlation coefficient with strength, all top models include the vacancy migration energy (VME). Beyond that, many combinations of other factor groups (e.g., surface energy, lattice/elastic constants) are essentially equivalent, with very similar root mean square error (RMSE) values as shown by the color map at the top of the figure. The top regression models perform similarly well with an average RMSE of approximately 200 MPa, or about 10\% of simulated plastic strength averaged over all 153 sampled IPs. \Cref{fig:linear_3factors} presents one of the best 3-variable regression models (see the figure caption for details), fit to the 153 samples remaining in the statistical pool with the adjusted goodness of fit parameter $r^2$ = 0.88.

\section*{Predicting flow strength by regression on ground-truth}

\begin{figure}
    \centering
    \includegraphics[width=3in]{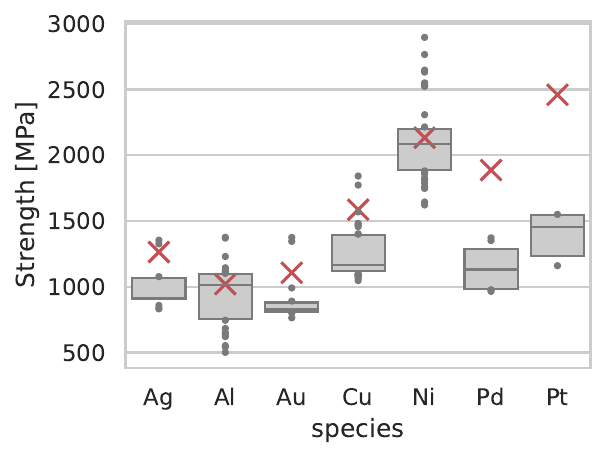}
    \caption{Plastic strength of seven metals predicted using a multi-linear regression model trained on three predictors for which DFT data was computed (\{111\} surface energy, lattice constant, and vacancy migration energy, all for FCC). Strength predicted from the DFT predictor values is shown using a red X.
    Gray boxes show quartiles of flow strength values observed for each metal in large-scale MD simulations for the 153 IPs included in this study. Gray circles indicate flow strength simulation results that are outside of the quartiles.}
    \label{fig:dft}
\end{figure}

We used GT DFT calculations to compute the predictor properties used in the 3-variable regression model shown in \Cref{fig:linear_3factors}, i.e., the \{111\} surface energy, the lattice constant, and the VME. \Cref{fig:dft} shows the flow strength of seven FCC metals Ag, Al, Au, Cu, Ni, Pd and Pt, predicted by substituting into the regression model the three predictors computed for each metal using DFT (see the Methods section for the details of the DFT calculations and the SI for a table of DFT results). Assuming that the 153 IPs and the seven DFT models belong to the same statistical pool (see explanation above), these values represent what seven hypothetical MD simulations with DFT forces would have predicted if indeed they could be performed under the same simulation conditions as those performed with the IPs. Under the same assumption, the pool of 153 IP samples used for constructing the regression model can also be used to estimate the uncertainty in the model’s predictions, e.g., using leave-one-out cross validation \cite{stone1974cross}.
Here, we fit 153 regression models on the same three predictors, leaving out one of the 153 IP samples from each model for validation. As an estimate of the prediction error by regression on DFT predictors, the average relative prediction error among 153 leave-one-out models is 15\%.
Five of seven DFT predictions fall within the 25-75\% inter-quartile range of MD flow strength values computed using the IPs, when the prediction error ($\pm 2\times\text{RMSE}$) is considered.
It is notable, however, that the predictions for Pd and Pt, even when accounting for their error bars (not shown), do not fall within the inter-quartile range, whereas they do for the other elements.
This is likely because the DFT values of the dominant VME predictor computed for these metals fall well outside the inter-quartile range of VME values of the 153 IPs (see SI). Likely informative, the source of this disagreement remains to be established.

\section*{Outlook}

In this work, we propose and demonstrate a statistical approach of cross-scale covariance in which a large-scale QoI is predicted by regression on small-scale properties of the same material model.
Our study of statistical covariance places on quantitative footing the long standing approach in atomistic simulations in which an IP is fitted to a set of small-scale properties deemed by an expert to be ``important'' for a particular material property or behavior of interest.
While the cross-scale relationships considered here have been assumed to exist in a broad context of materials research, here we establish and quantify covariance of plastic flow strength on a variety of small-scale properties through a data-driven approach, without an \textit{a priori} bias.

The study presented here is proof-of-principle and likely to suffer shortcomings not anticipated at its inception. Well-known in other contexts where statistical studies are common are issues of insufficient size and diversity of the statistical sample. In addition to having to pare down our statistical pool from 178 to 163 to 153 IP models, our statistical sample is likely to be still poorer than that. Among the 153 IPs remaining in the pool some may be ``clones'' in the sense that both the large-scale strength predictions and small-scale indicator properties computed with these potentials are very close. Yet to be verified, this is supported by the fact that many IPs suspected to be clones come from related groups of authors. Another issue is that nearly one third of all IP models included in our statistical pool were developed for nickel, whereas some other metals are represented by only few IP models. Further, some of the IPs included in the study are directly fitted to or post-validated against the properties used as predictors in our regression model, which biases the statistics by further reducing the diversity of our statistical pool of IP models.

Leaving aside known and as yet unknown shortcomings of our exploratory study, the goodness of fit parameters $r^2$ of some of our multi-linear regression models are approaching 0.9, whereas any value above 0.7 is commonly viewed as high. Is it possible to establish a still tighter covariance of plastic strength on small-scale properties? Two examples described above suggest that it may indeed be possible. In one, by focusing on regression outliers, we discovered SF jamming as an unusual new mechanism of resistance to dislocation motion in some IP models. This led us to exclude — prior to statistical analyses — all such models from our statistical pool, which improved the covariance. To our knowledge, SF jamming has not been observed before and may deserve further attention. In another example, physical insights into mechanisms of dislocation motion suggested additional indicators related to vacancy defect energetics not previously included in our covariance study. Including these new predictors improved the regression. These two examples illustrate how physical insights and statistical analyses can mutually benefit each other.

In summary, we propose an approach in which large-scale QoIs from MD simulations using an IP can be predicted by statistical regression on small-scale properties of the same IP model.
Taking advantage of the already large and rapidly increasing number of IP models and property calculations available and documented in the OpenKIM repository, along with the recently developed capability for large-scale MD simulations of plastic flow strength, here we establish a close covariance of the MD computed plastic flow strength with small-scale predictor properties of FCC metals.
The resulting cross-scale covariance relates small-scale properties, requiring only on the order of $10^4$ atoms$\times$steps or less to compute, to a large-scale QoI requiring $10^{14}$ atoms$\times$steps to simulate, thus bridging a scale gap of some 10 orders of magnitude.
The selection of which predictor properties to use is data-driven, accounting for statistical collinearity with the QoI and between the properties themselves, as well as the relative cost and robustness of GT computational methods.
The frequency of appearance in the top regression models served as further justification for the factor selection, allowing a straightforward method that can be used in property selection.
Our statistical analyses reveal several predictor properties most covariant with metal strength that can be recommended for inclusion in fitting or validation of bespoke IPs intended specifically for accurate simulations of plastic strength.
In addition to the regression models themselves, our statistical analyses estimate prediction uncertainties, making it meaningful to use resulting regression models to predict, within estimated uncertainty bounds, flow strength of crystals from GT theory.

\section*{Methods}
\label{sec:methods}

\subsection*{OpenKIM}
\label{sec:openkim}

OpenKIM\cite{TES11,tadmor2013nsf} (Open Knowledgebase of Interatomic Models) is a multifaceted cyberinfrastructure project founded in 2009. All content on OpenKIM is publicly available at \url{https://openkim.org}. The components of OpenKIM enabling the present work are the IP repository, the KIM API, and the automatic property testing framework.

At the time of submission, the OpenKIM IP repository contains 643 curated IPs of various types. KIM IPs conform to the KIM API that allows them to be used in plug-and-play fashion with compatible simulations platforms, such as LAMMPS \cite{thompson2022lammps}, ASE \cite{ase}, and others. IPs are gathered from literature, group websites, or other repositories such as the NIST IPR \cite{nist_ipr} by OpenKIM staff or submitted through the OpenKIM website by authors.

Every IP uploaded to OpenKIM is automatically run through a battery of property calculations using robust, vetted computational protocols (called ``KIM Test Drivers'') by the OpenKIM Processing Pipeline \cite{karls2020openkim, karls:clark:2022}. The pipeline is a cloud-based engine that matches IPs to compatible calculations (e.g., matching species and simulation code) and automatically runs the calculations on available HPC resources. Verification Checks (VCs) ensure an IP's coding and basic physics (e.g., invariances and smoothness). A KIM Test computes an IP's prediction for a material property and reports it in a standardized format conforming to a KIM Property Definition schema \cite{kimprops}.
The OpenKIM pipeline stores all results in a public, queryable MongoDB instance, which facilitates analyses across large samples of IPs such as the one done here and in~\cite{watersAutomatedDeterminationGrain2023}.

All non-DFT small-scale property calculations in the present work were performed by the OpenKIM Processing Pipeline. The flow strength calculations were facilitated by the KIM API, which enabled batch creation of LAMMPS input scripts with a uniform syntax independent of potential type. Because of the extreme scale of the plastic flow strength calculations, many of the KIM IPs had to be converted to an alternate (KIM Simulator Model) format that was compatible with the Kokkos acceleration package in LAMMPS.

\subsection*{MD simulations of plastic strength}
The LAMMPS code \cite{thompson2022lammps} was used to compute the plastic flow strength of 178 model materials, using large-scale MD simulations similar to those previously described in \cite{zepeda2017probing,zepeda2021atomistic}. First, a 3D periodic FCC single crystal with 54 million atoms was constructed in an initially orthorhombic supercell with an aspect ratio close to 1.23:1:0.81 and with edges oriented along the $[110]$, $[001]$ and $[1\bar10]$ directions of the cubic lattice. To seed dislocation sources, rhomb-shaped prismatic loops of the vacancy type were inserted precisely as described in \cite{zepeda2021atomistic}.  MD simulations were performed in an NVT ensemble at a temperature of 300~K under compression along the $x$-axis at a constant true strain rate of $10^8$/s.  To maintain a constant volume, the crystal was forced to expand at the same true rate along the $z$-axis, while its dimension along the $y$-axis was maintained constant. In the literature such specific straining geometry is often referred to as plane strain or channel-die compression \cite{basson2000deformation}. Before starting a simulation, the crystal's dimensions were scaled to match the lattice constant of each simulated FCC metal and equilibrated at 300~K and zero pressure. Among the 178 model materials, Poisson's ratio varied in the range 0.3--0.35, resulting in non-zero diagonal stress components along three axes of the specimen.
To account for all three non-zero diagonal components of the flow stress tensor, plastic strength was computed as the von Mises stress invariant \cite{timoshenko1983history}.

Out of the many IPs available and documented in the OpenKIM repository, 178 IPs previously developed for nine FCC metals were selected.
This included all 144 computationally expedient IPs of the embedded-atom model (EAM) \cite{daw1993embedded} and the Finnis-Sinclair (FS) \cite{finnis1984simple} types.
Furthermore, 34 more expensive IPs were included, listed in order of increasing computational cost: spectral neighbor analysis (SNAP) \cite{thompson2015spectral}, angular-dependent (ADP) \cite{mishin2006angular} and modified embedded-atom (MEAM) \cite{baskes1992modified} potentials. Although further diversification over IP types was desirable, being limited to our large but still finite HPC resource, we could not afford to run crystal strength simulations with still more expensive IPs available in OpenKIM. For the 34 moderately expensive potentials that we did use, we had to pare down the size of our simulated crystals to about 12 million atoms. As was documented in \cite{zepeda2021atomistic}, further reduction in crystal size results in increasing amplitude of thermal noise, thus raising measurement uncertainty in the predicted strength values. Given the size of our simulated crystals and about one million time steps needed to complete a simulation of crystal strength, we estimate to have executed in this series of simulations over $2 \times 10^{16}$  atoms$\times$steps.

\subsection*{DFT calculations of predictors}
Density functional theory (DFT) calculations were performed using the PBEsol exchange-correlation functional and pseudo-potentials based on the projector augmented wave (PAW) method \cite{PhysRevB.50.17953} available in the Vienna Ab initio Simulation Package (VASP) \cite{KRESSE199615}.
To accurately account for the inner-core and valence electrons, pseudopotentials for Al, Ag, Au, Cu, Ni, Pb, Pd, Pt, containing 3, 11, 11, 11, 10, 4, 10 and 10 valence electrons, respectively, were used.
Electronic structure calculations were performed by using a plane-wave energy cutoff of 600 eV and Methfessel-Paxton smearing scheme with a width of 0.15 eV.
Brillouin zone integrations were performed by using 32$\times$32$\times$32 Monkhorst-Pack mesh sizes for conventional FCC unit cells (each containing four atoms) or its equivalent for larger supercells.
Structural optimizations were terminated when total energies were converged to less than 10$^{-7}$ eV.
With this convergence criterion, atomic forces were typically converged to less than 0.001 eV/$\rm\AA$.

Elastic constant calculations were performed by using the procedure described in Ref.~\cite{PhysRevB.48.5844} with conventional unit cells containing four atoms.
To this end, we used strains between -2 to 2 \% and total energies for different strains were fitted to a sixth-order polynomial.
Vacancy formation energies were calculated using 4$\times$4$\times$4 supercells containing 255 atoms and one vacancy (with lattice vectors aligned with the three Cartesian coordinates) and $8\times8\times8$ {\bf k}-point meshes (equivalent to $32\times32\times32$ mesh for a unit cell), but due to memory limitations, vacancy migration barriers were calculated using $6\times6\times6$ {\bf k}-point meshes.

For surface energy and stacking fault energy calculations, supercells containing 224 atoms with lattice vectors parallel to $[111]$, $[01\bar{1}]$ and $[\bar{2}11]$ directions were used.
Each supercell consisted of 28 $(111)$ atomic layers, and free surface condition was imposed along $[111]$ direction and periodic boundary conditions were imposed along the two orthogonal directions.
Consequently, the edge length along $[111]$ was $>$80 $\rm\AA$ (thickness of vacuum region was set to 25\% of edge length along $[111]$) and the edge lengths along the two other orthogonal axes were typically less than 10 $\rm\AA$.

\section*{Data availability}
The python code and supporting data used for this paper will be made available on GitHub at \url{https://github.com/bjasperson/strength_covariance}.
Data pertaining to large-scale MD simulations are available on reasonable request.
All KIM test results, cited in the Supplementary Information and used to generate the indicator properties, are available at \url{https://openkim.org/}.

\section*{Author contributions}
VVB and EBT developed the concept, VVB and IN planned and performed MD simulations, BAJ and FZ performed statistical analyses, AS performed DFT calculations, BAJ and IN collected small-scale property data, VL secured funding support and managed the project.  All authors analyzed the findings and wrote the paper.

\section*{Competing interests}
The authors declare no competing interests.

\section*{Acknowledgements}

The authors acknowledge all developers of OpenKIM content (see SI for citations) and especially E.\ Fuemmeler for assisting in the development of the new vacancy property calculations.
The authors also acknowledge useful discussions with Y.\ Kurniawan and M.\ Transtrum.
BAJ also acknowledges useful discussions with H.\ T.\ Johnson.
BAJ acknowledges that this material is based in part upon work supported by the National Science Foundation (NSF) under Grant No.\ 1922758.
EBT and IN acknowledge partial support through NSF under Grants No.\ 1834251, 2333411 and 1834332.
All authors acknowledge funding support from the Laboratory Directed Research and Development program (tracking number 23-SI-006) and a special computational time allocation on Lassen supercomputer from the Computational Grand Challenge program at Lawrence Livermore National Laboratory.
This work was performed under the auspices of the U.S.\ Department of Energy by Lawrence Livermore National Laboratory under Contract DE-AC52-07NA27344.

\begin{singlespace}
	\printbibliography[heading=subbibintoc]
\end{singlespace}

\end{refsection}


\newpage
\section*{Supplementary Information for ``Cross-scale covariance for material property predictions''}
  
\newcommand{\beginsupplement}{%
	\setcounter{table}{0}
	\renewcommand{\thetable}{S\arabic{table}}%
	\setcounter{figure}{0}
	\renewcommand{\thefigure}{S\arabic{figure}}%
	\setcounter{section}{0}
	\renewcommand{\thesection}{S\arabic{section}}%
}
\beginsupplement 
\begin{refcontext}{supp}

\section{Property outlier cutoffs for exclusion from dataset}
Property tests are automatically run for relevant models uploaded to OpenKIM.
There exist a number of reasons why a property calculation may result in an outlier value for a given model.
For example, property predictions may be generated automatically by the KIM pipeline for crystal structures that are not intended in the original model fitting.
As a result, it is inevitable that some property calculations will result in what can reasonably be considered outlier values. 
The list below outlines the outlier cutoff values used in generating the training dataset for model development:
\begin{enumerate}
	\item All bulk modulus values greater than 100,000 GPa
	\item All elastic constants greater than 60,000 GPa
	\item All cohesive energy predictions greater than 45 eV
	\item All stacking energies greater than 0.45 eV/A\textsuperscript{2}
\end{enumerate}

\section{Detailed correlation heatmap and covariance plots}

\Cref{fig:corr_plot_full} shows a full correlation heatmap of all available properties with strength. 
Values of -1 and 1 are strongly correlated. 
The square regions along the diagonal indicate strongly correlated indicator properties, e.g. surface energies and lattice constants.

Additional covariance plots are shown in \Cref{fig:pp_jammed,fig:pp_wo_jammed,fig:pp_se,fig:pp_stack_twin}.
\Cref{fig:pp_jammed,fig:pp_wo_jammed} highlight select indicator properties before and after removing IPs that exhibited ``jamming'' (see Main text for details).
Strong correlation between potential indicator properties are shown in \Cref{fig:pp_se,fig:pp_stack_twin}.

\begin{figure}
	\centering
	\includegraphics[]{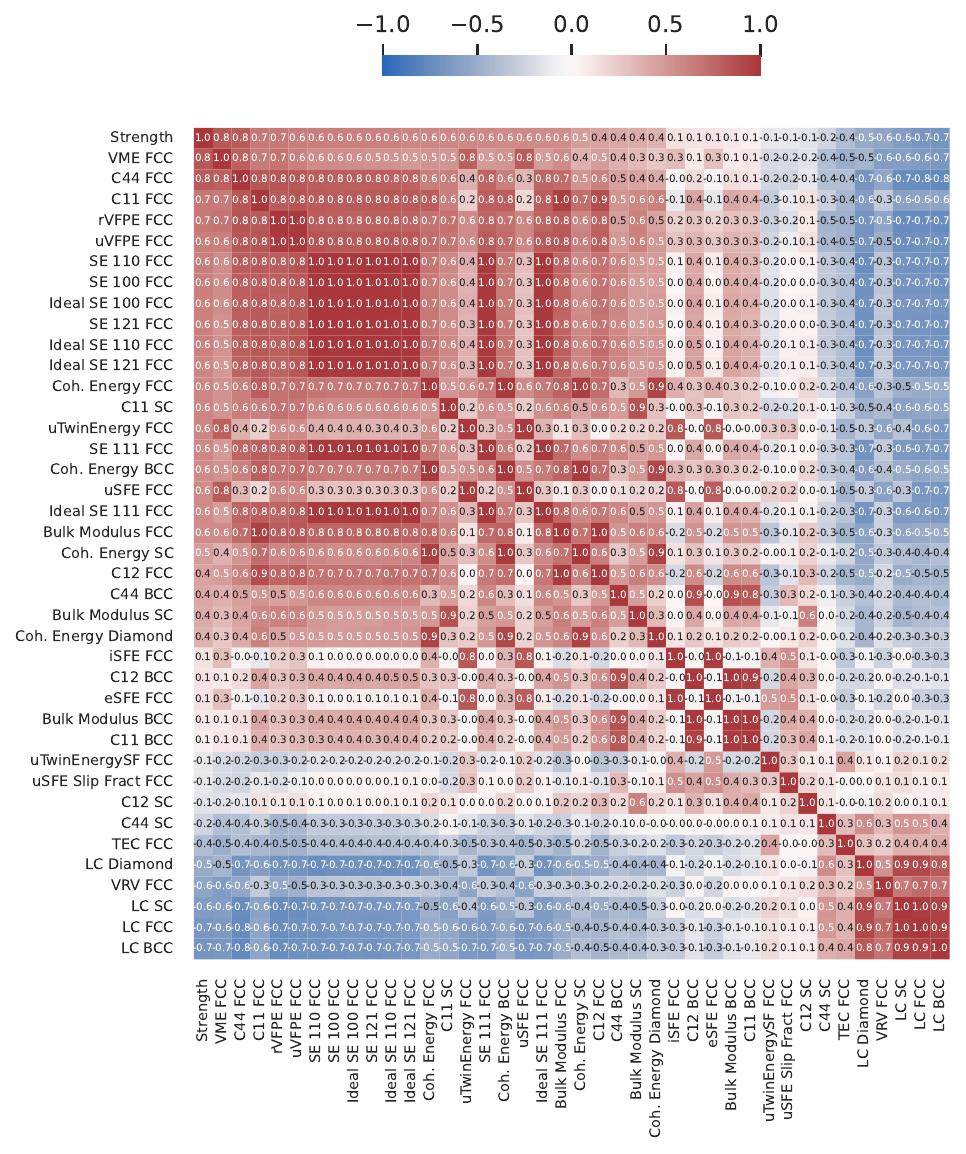}
	\caption{Full correlation heatmap of all properties with strength, sorted by correlation coefficient with strength. Values of -1 and 1 are strongly correlated. The square regions along the diagonal indicate strongly correlated indicator properties, including surface energy and lattice constants. VME = vacancy migration energy, rVFPE/uVFPE = relaxed/unrelaxed vacancy formation potential energy, SE = surface energy, Coh. Energy = cohesive energy, uTwinEnergy = unstable twinning energy, iSFE/eSFE/uSFE = intrinsic/extrinsic/unstable stacking fault energy, TEC = thermal expansion coefficient, LC = lattice constant, VRV = vacancy relaxation volume.}
	\label{fig:corr_plot_full}
\end{figure}

\begin{figure}
	\centering
	\includegraphics[width=6in]{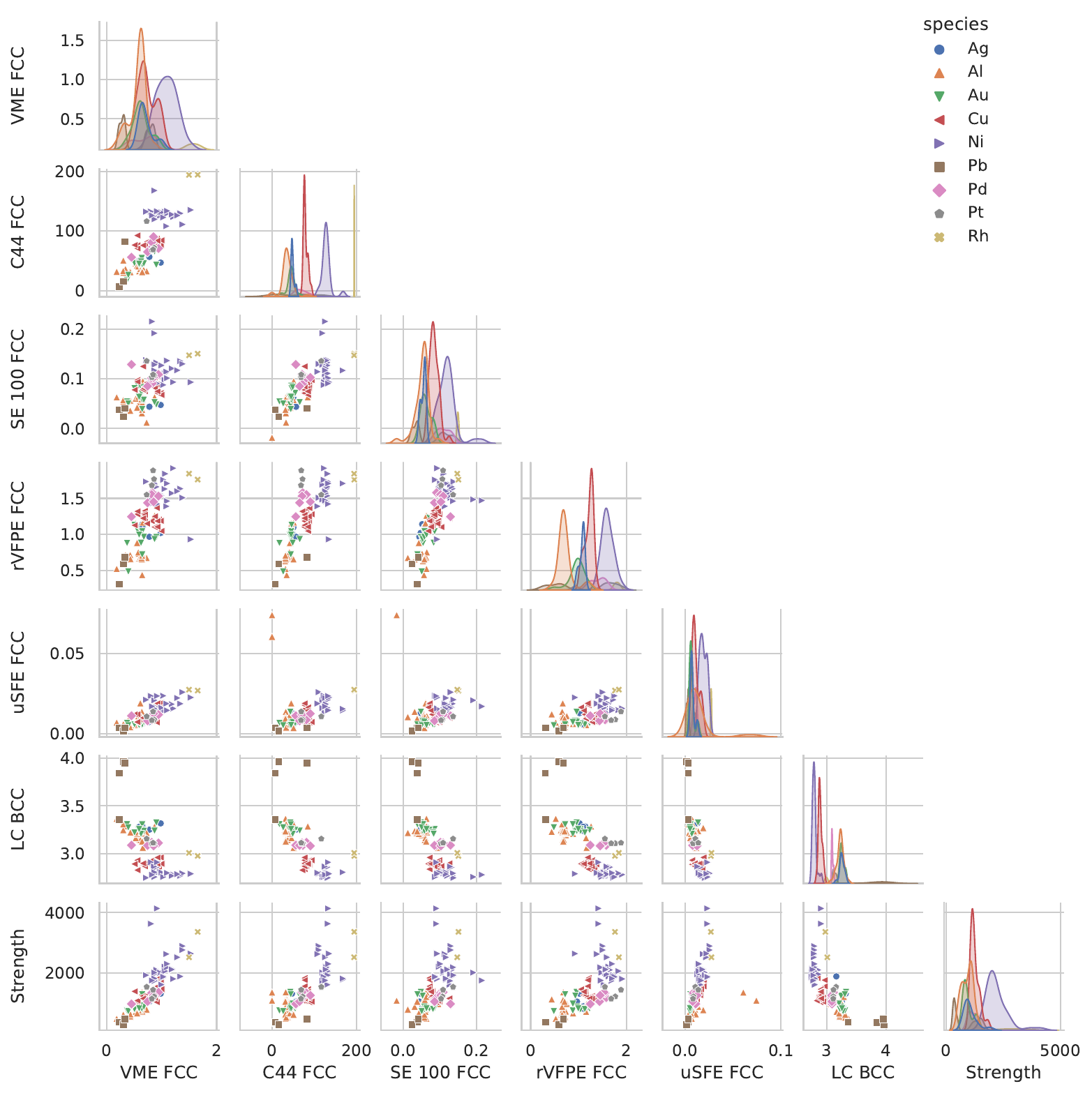}
	\caption{Scatter plots across select potential indicator properties and strength, prior to removing IPs that exhibited ``jamming''. Properties include vacancy migration energy, $c_{44}$ shear modulus, surface energy {100}, relaxed vacancy formation potential energy, unstable stacking fault energy, and lattice constant.}
	\label{fig:pp_jammed}
\end{figure}

\begin{figure}
	\centering
	\includegraphics[width=6in]{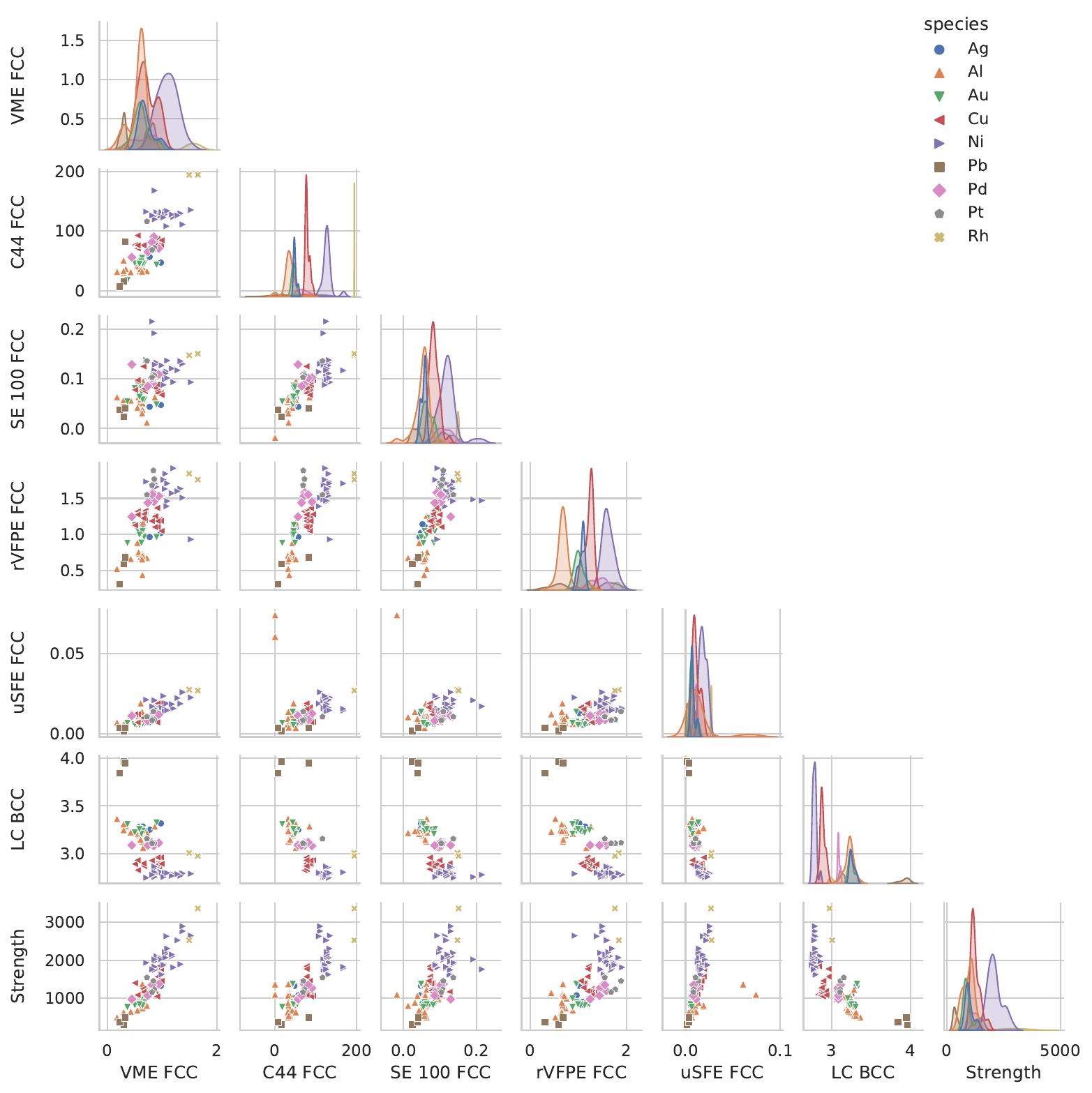}
	\caption{Scatter plots across select potential indicator properties and strength, after removing IPs that exhibited ``jamming''. Properties include vacancy migration energy, $c_{44}$ shear modulus, surface energy {100}, relaxed vacancy formation potential energy, unstable stacking fault energy, and lattice constant.}
	\label{fig:pp_wo_jammed}
\end{figure}

\begin{figure}
	\centering
	\includegraphics[width=6in]{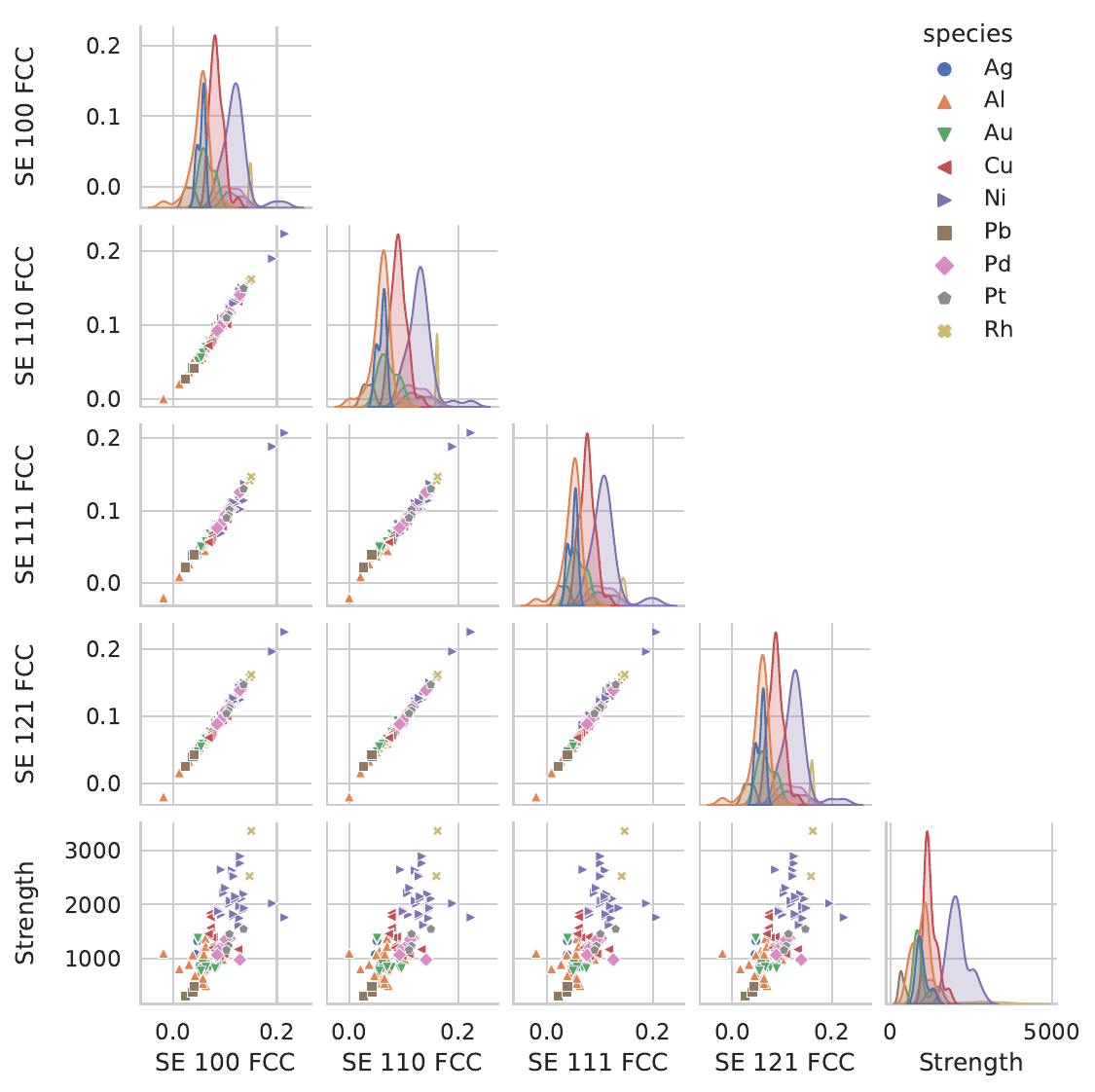}
	\caption{Scatter plots of strength and surface energies across various surfaces, defined by the Miller indices of the crystal surface. Strong cross-correlation between surface energies is observed.}
	\label{fig:pp_se}
\end{figure}

\begin{figure}
	\centering
	\includegraphics[width=6in]{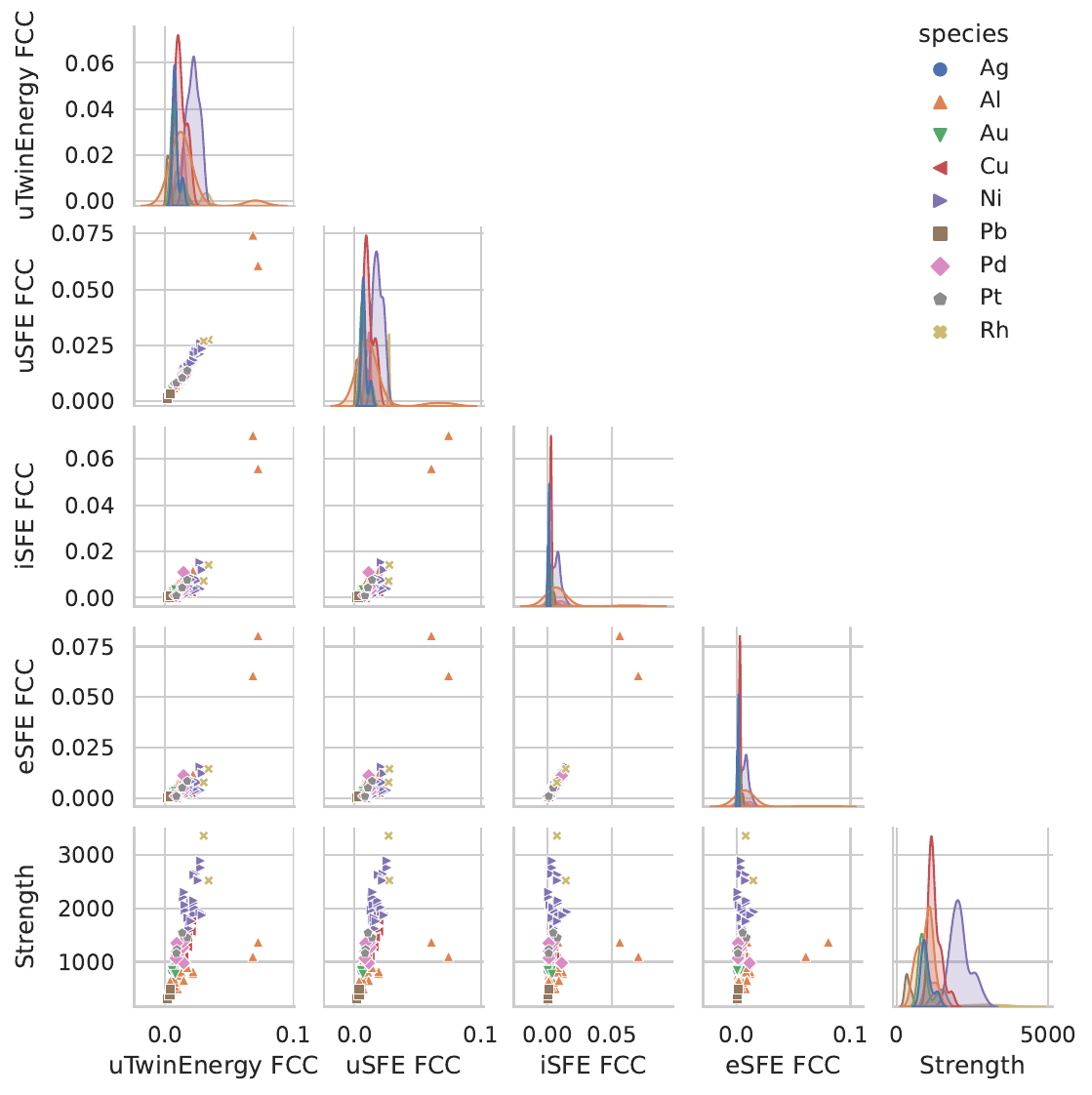}
	\caption{Scatter plots of strength and various stacking energies, including unstable twinning energy and unstable/intrinsic/extrinsic stacking fault energies. Of note are the strong cross-correlation between uSFE/uTwinEnergy and eSFE/iSFE}
	\label{fig:pp_stack_twin}
\end{figure}

\section{Data imputation for missing property values}
Estimators such as linear regression and Support Vector Regression in \textsc{sklearn} require no missing values. 
Every effort was made to simulate all canonical properties for each model used in the training data. 
Nevertheless, there were some cases where this was not feasible for all models across all properties explored.
Rather than discard these models from the analysis, a commonly used method called data imputation was implemented to fill in the gaps.

The approach used, KNNImputer, finds the nearest neighbors (in this case, set to 2) based on Euclidean distance.
Since the dataset contained missing values (hence the need for data imputation), a method of computing distances in the presence of these missing values was needed.
Here we used the \textit{nan euclidean distances} approach, which uniformly weighs the coordinates that are present. 
See the corresponding \href{https://scikit-learn.org/stable/modules/generated/sklearn.metrics.pairwise.nan_euclidean_distances.html#sklearn-metrics-pairwise-nan-euclidean-distances}{user guide} for details.

\section{Additional regression and model analysis}
The results reported in the main text are based on linear regression models of the indicator properties. To test whether and to what extent nonlinear models might be able to better fit the observed data, we test two additional regression methods: polynomial regression and symbolic regression (SR). One has to be careful about any nonlinear regression models as there are only 163 data points, including some with certain indicator properties missing. It is easy to over-fit such a small dataset with nonlinear models. Indeed, if we include 163 or more degrees of freedom in a fit, a number easy to exceed since nonlinear regressions can transform or combine the original input variables into many effective degrees of freedom, one can reduce the fitting error to zero but with no predictive power.

\subsection{Symbolic regression}
Symbolic regression (SR) is a nonlinear regression method that can be used for fitting data and discovering physical rules \cite{Schmidt2009S-Distilling}.  SR implementations typically rely on Monte-Carlo based algorithms to search for optimal expressions, and results are usually the best local minimum solution within a given search budget, rather than the global minimum. As a result, SR is often orders of magnitude more expensive than linear or polynomial regression and hard to scale to many variables.  Therefore, we employ it sparingly for finding a closed-form expression using only a small number of indicators. In this work, we iterate over all size 1, 2 or 3 subsets of the 27 indicator variables. 
These variables are: lattice constant (fcc, bcc, sc), bulk modulus (fcc, bcc, sc), c\textsubscript{11}/c\textsubscript{12}/c\textsubscript{44} (fcc, bcc, sc), cohesive energy (fcc, bcc, sc), thermal expansion coeffient (fcc), surface energy (\{100\}, \{110\}, \{111\}, \{121\}) (fcc), extrinsic/intrinsic/unstable stacking fault energy (fcc), and unstable twinning energy (fcc). The ``DeepSymbolicRegressor'' class within the Deep Symbolic Optimization (DSO) package \cite{petersen2021deep} was adopted. The mathematical expressions considered were the basic arithmetic operators as well as ``constant'', ``exp'', ``log'', and ``poly''.  For each selected set of variables, the search is limited to $10^4$ samples with batch size of $10^3$ and 2 CPU-cores per batch. For the best SR fits, extended search of $10^5$ samples was performed, but no significantly better fits were discovered. 
\Cref{fig:SR-single} shows two examples of single-variable symbolic regression fits.
\begin{figure}
	\centering
	\includegraphics[width=0.45\textwidth]{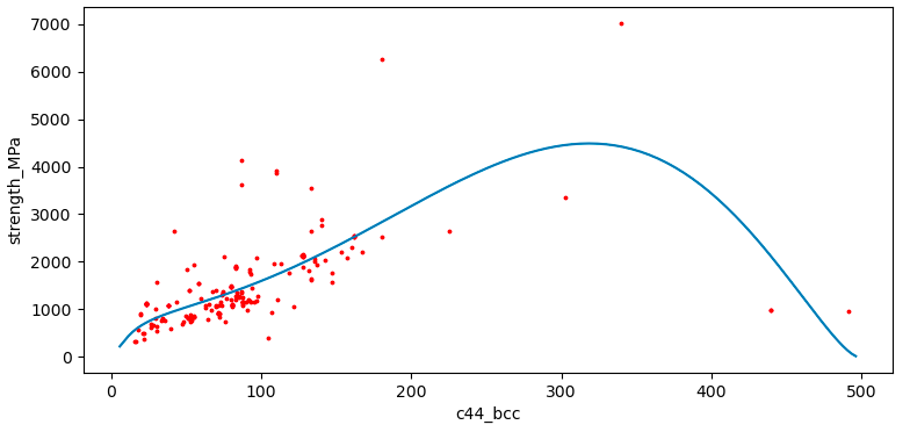}
	\includegraphics[width=0.45\textwidth]{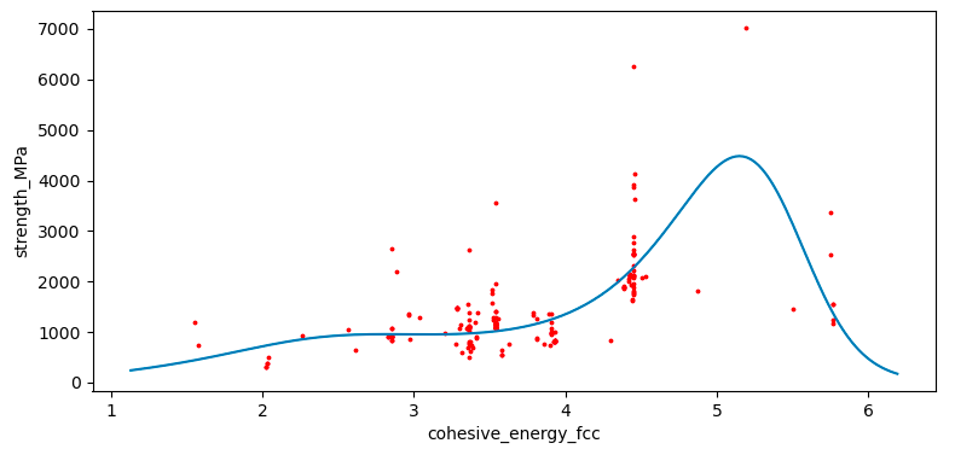}
	\caption{Examples of symbolic regression fitting of strength using a single variable. a) $c_{44}$ of bcc and b) cohesive energy of fcc.}
	\label{fig:SR-single}
\end{figure}

\begin{table}[]
	\centering
	\caption{Examples of best symbolic regression $R^2$ values using 2 or 3 variables.}
	\begin{tabular}{llll}
		\hline
		variables & $R^2$ (linear) & $R^2$ (SR) &  SR - linear \\ \hline
		SE 111 FCC; uSFE fcc		&0.686	&0.797	&0.111 \\
		SE 110 FCC; uSFE fcc		&0.690	&0.811	&0.121 \\
		\hline
		c44 sc; SE 111 FCC; uSFE FCC			&	0.714	&    0.904	  & 0.190\\
		SE 111 fcc; uSFE fcc; uTwinEnergy fcc	& 0.715	  & 0.913	    & 0.197 \\ 
	\end{tabular}
	\label{tab:SR}
\end{table}

\Cref{tab:SR} shows the best symbolic regression fits using either 2 or 3 variables. Overall, the identified variables are consistent with those from linear fitting. However, naive $R^2$ from a SR fitting is likely too optimistic, and obtaining more predictive metrics of fitting using e.g.~ leave-$k$-out cross validation is computationally too slow. We leave more detailed analysis of nonlinear fitting to polynomial regression in the next subsection. 

\subsection{Polynomial regression}
In polynomial regression (PR), the $M$ indicator properties $\{x_1, \dots, x_M\}$, as well as the polynomials up to order $N$
are used for a linear regression. Note that a polynomial regression provides a nonlinear model to the data, but the regression itself is linear in the unknown coefficients.
This makes PR much more manageable than SR. 

Since the total number of data points (up to 163) is small, three measures were taken to prevent overfitting. First, we chose $N=2$, i.e. quadratic polynomials ($x_i^2$ and cross terms $x_i x_j$) in order to limit the number of terms in the fitting. Second, we employed leave-one-out cross validation error as the metric of regression quality, rather than the fitting errors (MSE, MAE, etc). In leave-$k$-out cross validation, $k$ data points are excluded from the training set, the prediction errors over the $k$ left-out points using the fit from $N-k$ points are recorded, and the obtained errors are averaged over all $k$-subsets of the data set. Leave-one-out is the simplest approach, with $k=1$, and the average prediction error over $N$ data points are considered as a figure of merit of the polynomial regression. Finally, a systematic search of important or predictive indicator variables was performed: we loop over all possible $m$-subsets of the $M=27$ indicator properties in order to reduce the number of variables considered in each fitting and suppress overfitting. For example, all possible pairs of indicators such as \{surface energy, stacking fault energy\} were considered. Adopting polynomial regressions, we examined all singlets, pairs and triplets of indicator variable combinations as well. Each fit has $(m+1)(m+2)/2$ unknown coefficients. 

The results of leave-one-out cross validation are shown in \Cref{fig:PR}. In panel a), the polynomial fits (y-axis) are compared to linear models (x-axis) and shown to increase the cross validation  in most cases. In the best models (top right corner), the improvements of nonlinear polynoial fits are the most modest. In panel b), the fits are listed from single variable (left of divider), pair of variables and triplet (rightmost). The triplet models using 3 variables are the most accurate, but the benefit of non-linearity is limited.
\Cref{fig:PR} shows two example single-variable symbolic regression fits.
\begin{figure}
	\centering
	\includegraphics[width=0.9\textwidth]{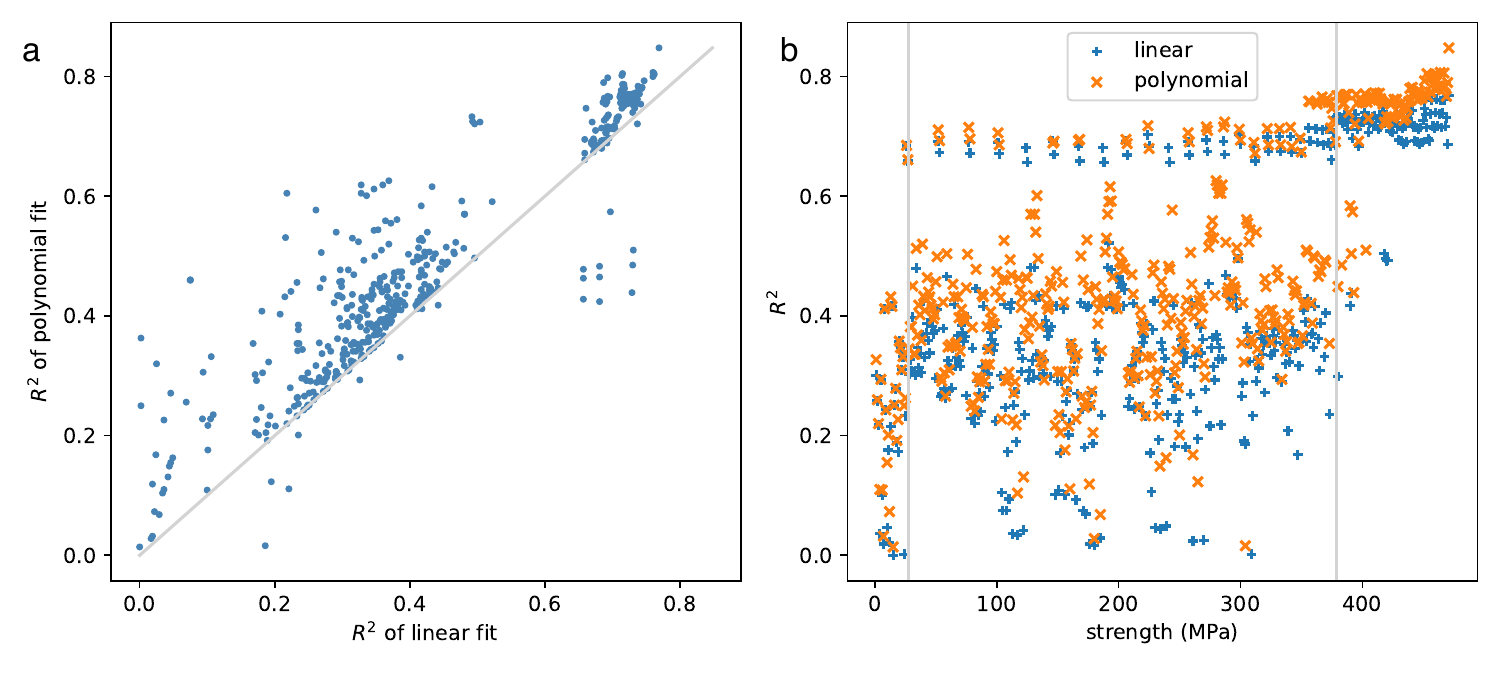}
	\caption{Results of leave-one-out cross validation of polynomial regression fitting of strength using up to 3 variables. a) Compared to linear fitting b) Listed with all attempted fits, where the vertical lines divide the single, pair and triplet models.}
	\label{fig:PR}
\end{figure}

In summary, nonlinear regressions were performed conservatively on small subsets of indicator properties. The results are in general agreement with the linear regression and provide at most marginally tighter fits than linear regression. A likely reason is that, due to the small dataset, we did not aggressively push for complicated or highly nonlinear fits.

\section{Top regression model predictor groups}
Different properties are combined into ``property groups'' to aid in identifying property significance in the top performing regression models, see Figure 4 in the main manuscript. Rows that have multiple properties in their group include the following:
\begin{enumerate}
	\item elastic constants: $c_{11}$, $c_{12}$, $c_{44}$, bulk modulus
	\item surface energy: surface energy and ideal surface energy (\{100\}, \{110\}, \{111\}, \{121\}) 
	\item stack/twin fault energy: intrinsic and extrinsic stacking fault energy, unstable twinning energy and slip fraction,  unstable stacking fault energy and slip fraction
	\item vacancy formation energies: relaxed and unrelaxed vacancy formation potential energy
\end{enumerate}

\section{Density Functional Theory calculations and results}

\begin{table}[]
	\centering
	\caption{Calculated DFT properties across species (FCC crystal structure). See Methods section of main text for calculation details.}
	\begin{tabular}{lllllll}
		\hline
		Species & LC (A)  & c44 (GPa)  & SE 111 (J/m\textsuperscript{2}) & iSFE (J/m\textsuperscript{2})   & rVFPE (eV)  & VME (eV)    \\ \hline
		Ag      & 4.0509 & 52.4133  & 0.9588 & 0.0241 & 0.7100 & 0.7886 \\
		Al      & 4.0139 & 36.1276  & 0.9580 & 0.1256 & 0.6880 & 0.6314 \\
		Au      & 4.0784 & 34.5375  & 0.9695 & 0.0323 & 0.3966 & 0.6900 \\
		Cu      & 3.5628 & 93.0490  & 1.6116 & 0.0473 & 1.0629 & 0.8495 \\
		Ni      & 3.4530 & 131.3062 & 2.3591 & 0.1566 & 1.4669 & 1.0718 \\
		Pd      & 3.8664 & 73.3238  & 1.6955 & 0.1680 & 1.2718 & 1.0538 \\
		Pt      & 3.9129 & 75.1432  & 1.8429 & 0.3170 & 0.7153 & 1.3969 \\ \hline
	\end{tabular}%
	\label{tab:dft_values}
\end{table}

\Cref{tab:dft_values} presents the calculated indicator property values using DFT, see Methods in the main text for calculation details.
\Cref{fig:dft_c44,fig:dft_iSFE,fig:dft_SE111,fig:dft_lc,fig:dft_VME,fig:dft_rVFPE} show the DFT properties calculated, along with a boxplot of the property values calculated across the IP models used in this work.

\begin{figure}
	\centering
	\includegraphics{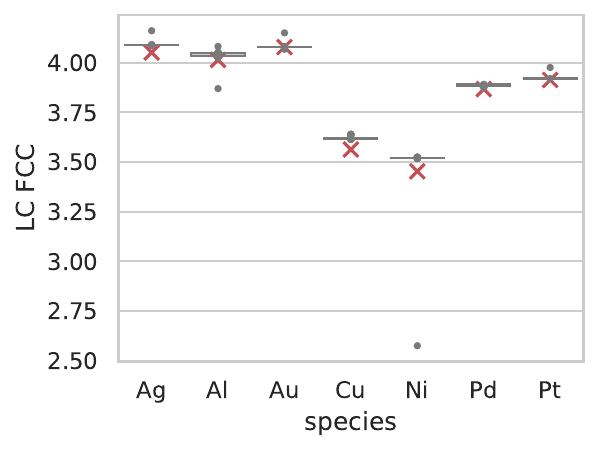}
	\caption{Lattice constant (Angstroms) across species, by prediction method. Red X's indicate calculated property value using DFT. Gray boxes show quartiles of the property values simulated for each metal amongst the 153 large-scale MD simulations included in this study. Gray circles indicate property simulation results that are outside of the quartiles.}
	\label{fig:dft_lc}
\end{figure}

\begin{figure}
	\centering
	\includegraphics{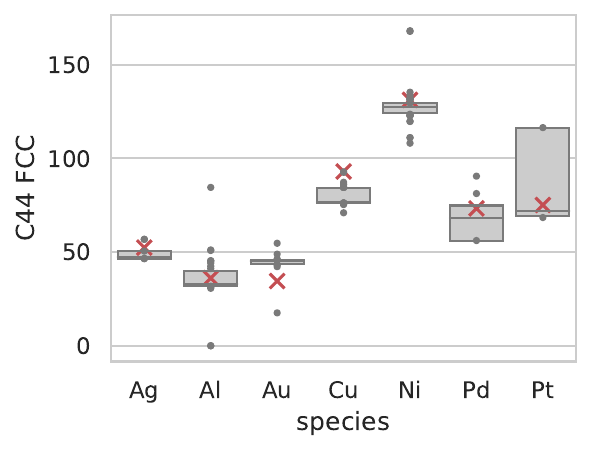}
	\caption{$C_{44}$ elastic constant (GPa) across species, by prediction method. Red X's indicate calculated property value using DFT. Gray boxes show quartiles of the property values simulated for each metal amongst the 153 large-scale MD simulations included in this study. Gray circles indicate property simulation results that are outside of the quartiles.}
	\label{fig:dft_c44}
\end{figure}

\begin{figure}
	\centering
	\includegraphics{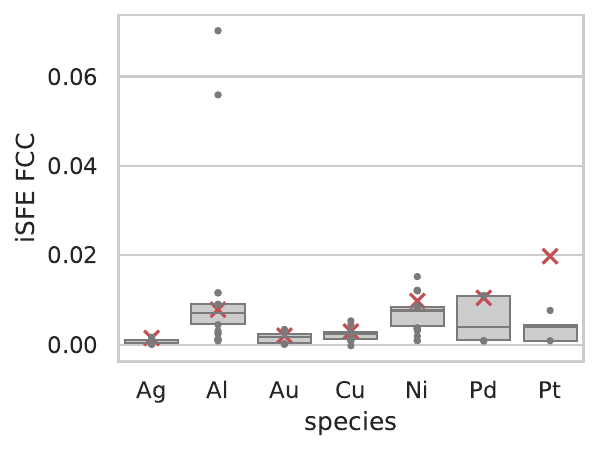}
	\caption{Intrinsic stacking fault energy (eV/A\textsuperscript{2}) across species, by prediction method. Red X's indicate calculated property value using DFT. Gray boxes show quartiles of the property values simulated for each metal amongst the 153 large-scale MD simulations included in this study. Gray circles indicate property simulation results that are outside of the quartiles.}
	\label{fig:dft_iSFE}
\end{figure}

\begin{figure}
	\centering
	\includegraphics{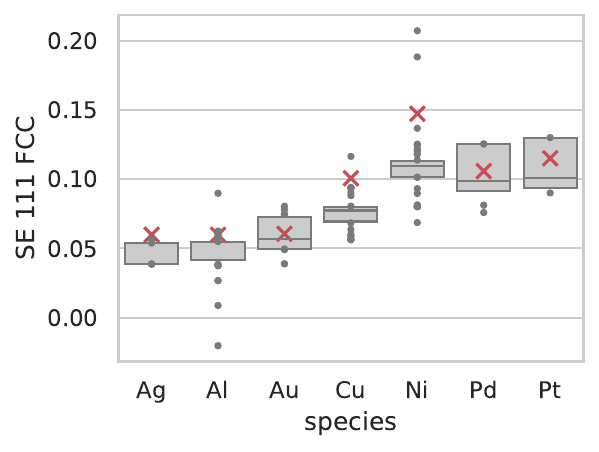}
	\caption{Surface energy {111} (eV/A\textsuperscript{2}) across species, by prediction method. Red X's indicate calculated property value using DFT. Gray boxes show quartiles of the property values simulated for each metal amongst the 153 large-scale MD simulations included in this study. Gray circles indicate property simulation results that are outside of the quartiles.}
	\label{fig:dft_SE111}
\end{figure}

\begin{figure}
	\centering
	\includegraphics{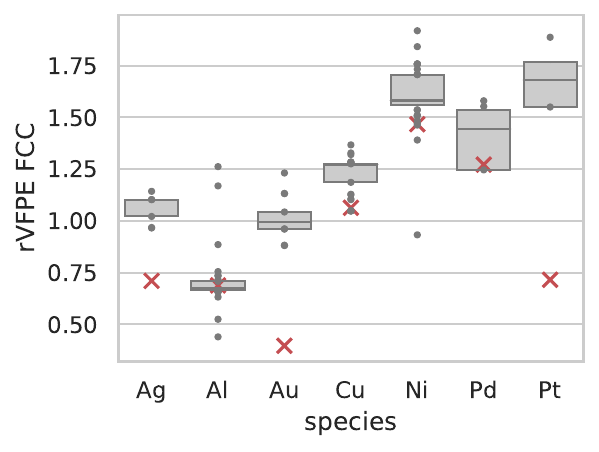}
	\caption{Relaxed vacancy formation potential energy (eV) across species, by prediction method. Red X's indicate calculated property value using DFT. Gray boxes show quartiles of the property values simulated for each metal amongst the 153 large-scale MD simulations included in this study. Gray circles indicate property simulation results that are outside of the quartiles.}
	\label{fig:dft_rVFPE}
\end{figure}

\begin{figure}
	\centering
	\includegraphics{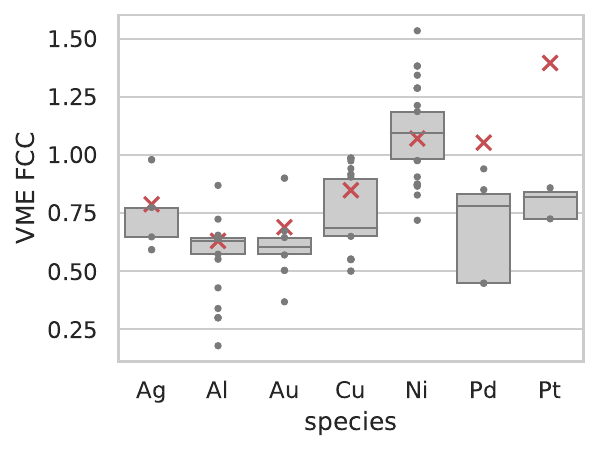}
	\caption{Vacancy migration energy (eV) across species, by prediction method. Red X's indicate calculated property value using DFT. Gray boxes show quartiles of the property values simulated for each metal amongst the 153 large-scale MD simulations included in this study. Gray circles indicate property simulation results that are outside of the quartiles.}
	\label{fig:dft_VME}
\end{figure}


\newpage
\section{Models and Test Drivers used}
The models and test drivers used in this work were originally published and also archived in OpenKIM through the following references: 

\begin{FlushLeft}
	\begin{multicols}{3}
		Ag, 832.0 \cite{OpenKIM-MO:318213562153:000a,OpenKIM-MO:318213562153:000,OpenKIM-MD:120291908751:005}\\ 
		Ag, 837.0 \cite{OpenKIM-MO:947112899505:005a,OpenKIM-MO:947112899505:005b,OpenKIM-MO:947112899505:005,OpenKIM-MD:120291908751:005}\\ 
		Ag, 858.0 \cite{OpenKIM-MO:222110751402:000a,OpenKIM-MO:222110751402:000,OpenKIM-MD:120291908751:005}\\ 
		Ag, 905.0 \cite{OpenKIM-MO:270337113239:005a,OpenKIM-MO:270337113239:005,OpenKIM-MD:120291908751:005}\\ 
		Ag, 908.0 \cite{OpenKIM-MO:128703483589:005a,OpenKIM-MO:128703483589:005,OpenKIM-MD:120291908751:005}\\ 
		Ag, 908.0 \cite{OpenKIM-MO:270337113239:005a,OpenKIM-MO:270337113239:005,OpenKIM-MD:120291908751:005}\\ 
		Ag, 912.0 \cite{OpenKIM-MO:104806802344:005a,OpenKIM-MO:104806802344:005b,OpenKIM-MO:104806802344:005,OpenKIM-MD:120291908751:005}\\ 
		Ag, 912.0 \cite{OpenKIM-MO:131620013077:005a,OpenKIM-MO:131620013077:005,OpenKIM-MD:120291908751:005}\\ 
		Ag, 923.0 \cite{OpenKIM-MO:108983864770:005a,OpenKIM-MO:108983864770:005b,OpenKIM-MO:108983864770:005,OpenKIM-MD:120291908751:005}\\ 
		Ag, 1064.0 \cite{OpenKIM-MO:681640899874:000a,OpenKIM-MO:681640899874:000,OpenKIM-MD:120291908751:005}\\ 
		Ag, 1077.0 \cite{OpenKIM-MO:626948998302:000a,OpenKIM-MO:626948998302:000b,OpenKIM-MO:626948998302:000,OpenKIM-MD:120291908751:005}\\ 
		Ag, 1326.0 \cite{OpenKIM-MO:055919219575:000a,OpenKIM-MO:055919219575:000,OpenKIM-MD:120291908751:005}\\ 
		Ag, 1354.0 \cite{OpenKIM-MO:212700056563:005a,OpenKIM-MO:212700056563:005,OpenKIM-MD:120291908751:005}\\ 
		Ag, 1895.0 \cite{OpenKIM-SM:485325656366:001a,OpenKIM-SM:485325656366:001}\\ 
		Al, 501.0 \cite{OpenKIM-MO:418978237058:005a,OpenKIM-MO:418978237058:005b,OpenKIM-MO:418978237058:005,OpenKIM-MD:120291908751:005}\\ 
		Al, 542.0 \cite{OpenKIM-MO:131650261510:005a,OpenKIM-MO:131650261510:005b,OpenKIM-MO:131650261510:005,OpenKIM-MD:120291908751:005}\\ 
		Al, 552.0 \cite{OpenKIM-MO:049243498555:000a,OpenKIM-MO:049243498555:000b,OpenKIM-MO:049243498555:000,OpenKIM-MD:120291908751:005}\\ 
		Al, 585.0 \cite{OpenKIM-MO:942551040047:005a,OpenKIM-MO:942551040047:005,OpenKIM-MD:120291908751:005}\\ 
		Al, 621.0 \cite{OpenKIM-MO:284963179498:005a,OpenKIM-MO:284963179498:005,OpenKIM-MD:120291908751:005}\\ 
		Al, 636.0 \cite{OpenKIM-MO:060567868558:000a,OpenKIM-MO:060567868558:000b,OpenKIM-MO:060567868558:000,OpenKIM-MD:120291908751:005}\\ 
		Al, 652.0 \cite{OpenKIM-MO:149316865608:005a,OpenKIM-MO:149316865608:005b,OpenKIM-MO:149316865608:005c,OpenKIM-MO:149316865608:005,OpenKIM-MD:120291908751:005}\\ 
		Al, 683.0 \cite{OpenKIM-MO:120808805541:005a,OpenKIM-MO:120808805541:005,OpenKIM-MD:120291908751:005}\\ 
		Al, 686.0 \cite{OpenKIM-SM:871795249052:000a,OpenKIM-SM:871795249052:000}\\ 
		Al, 731.0 \cite{OpenKIM-MO:411692133366:000a,OpenKIM-MO:411692133366:000,OpenKIM-MD:120291908751:005}\\ 
		Al, 745.0 \cite{OpenKIM-SM:113843830602:000a,OpenKIM-SM:113843830602:000}\\ 
		Al, 758.0 \cite{OpenKIM-MO:519613893196:000a,OpenKIM-MO:519613893196:000,OpenKIM-MD:120291908751:005}\\ 
		Al, 771.0 \cite{OpenKIM-MO:137572817842:000a,OpenKIM-MO:137572817842:000,OpenKIM-MD:120291908751:005}\\ 
		Al, 789.0 \cite{OpenKIM-SM:811588957187:000a,OpenKIM-SM:811588957187:000}\\ 
		Al, 809.0 \cite{OpenKIM-MO:577453891941:005a,OpenKIM-MO:577453891941:005,OpenKIM-MD:120291908751:005}\\ 
		Al, 817.0 \cite{OpenKIM-SM:721930391003:000a,OpenKIM-SM:721930391003:000}\\ 
		Al, 893.0 \cite{OpenKIM-MO:106969701023:005a,OpenKIM-MO:106969701023:005,OpenKIM-MD:120291908751:005}\\ 
		Al, 899.0 \cite{OpenKIM-MO:658278549784:005a,OpenKIM-MO:658278549784:005,OpenKIM-MD:120291908751:005}\\ 
		Al, 986.0 \cite{OpenKIM-MO:042691367780:000a,OpenKIM-MO:042691367780:000,OpenKIM-MD:120291908751:005}\\ 
		Al, 998.0 \cite{OpenKIM-MO:338600200739:000a,OpenKIM-MO:338600200739:000,OpenKIM-MD:120291908751:005}\\ 
		Al, 1027.0 \cite{OpenKIM-MO:123629422045:005a,OpenKIM-MO:123629422045:005,OpenKIM-MD:120291908751:005}\\ 
		Al, 1069.0 \cite{OpenKIM-MO:117656786760:005a,OpenKIM-MO:117656786760:005,OpenKIM-MD:120291908751:005}\\ 
		Al, 1075.0 \cite{OpenKIM-MO:101214310689:005a,OpenKIM-MO:101214310689:005,OpenKIM-MD:120291908751:005}\\ 
		Al, 1077.0 \cite{OpenKIM-MO:020851069572:000a,OpenKIM-MO:020851069572:000,OpenKIM-MD:120291908751:005}\\ 
		Al, 1077.0 \cite{OpenKIM-MO:664470114311:005a,OpenKIM-MO:664470114311:005,OpenKIM-MD:120291908751:005}\\ 
		Al, 1088.0 \cite{OpenKIM-MO:699137396381:005a,OpenKIM-MO:699137396381:005b,OpenKIM-MO:699137396381:005c,OpenKIM-MO:699137396381:005,OpenKIM-MD:120291908751:005}\\ 
		Al, 1096.0 \cite{OpenKIM-MO:678952612413:000a,OpenKIM-MO:678952612413:000,OpenKIM-MD:120291908751:005}\\ 
		Al, 1099.0 \cite{OpenKIM-MO:559870613549:000a,OpenKIM-MO:559870613549:000,OpenKIM-MD:120291908751:005}\\ 
		Al, 1102.0 \cite{OpenKIM-MO:651801486679:005a,OpenKIM-MO:651801486679:005,OpenKIM-MD:120291908751:005}\\ 
		Al, 1107.0 \cite{OpenKIM-SM:667696763561:000a,OpenKIM-SM:667696763561:000}\\ 
		Al, 1118.0 \cite{OpenKIM-MO:751354403791:005a,OpenKIM-MO:751354403791:005,OpenKIM-MD:120291908751:005}\\ 
		Al, 1126.0 \cite{OpenKIM-MO:826591359508:000a,OpenKIM-MO:826591359508:000,OpenKIM-MD:120291908751:005}\\ 
		Al, 1144.0 \cite{OpenKIM-MO:722733117926:000a,OpenKIM-MO:722733117926:000,OpenKIM-MD:120291908751:005}\\ 
		Al, 1230.0 \cite{OpenKIM-MO:109933561507:005a,OpenKIM-MO:109933561507:005,OpenKIM-MD:120291908751:005}\\ 
		Al, 1370.0 \cite{OpenKIM-MO:051157671505:000a,OpenKIM-MO:051157671505:000,OpenKIM-MD:120291908751:005}\\ 
		Al, 1376.0 \cite{OpenKIM-MO:019873715786:000a,OpenKIM-MO:019873715786:000,OpenKIM-MD:120291908751:005}\\ 
		Al, 1560.0 \cite{OpenKIM-SM:656517352485:000a,OpenKIM-SM:656517352485:000}\\ 
		Al, 2636.0 \cite{OpenKIM-MO:820335782779:000a,OpenKIM-MO:820335782779:000,OpenKIM-MD:120291908751:005}\\ 
		Au, 746.0 \cite{OpenKIM-SM:985135773293:000a,OpenKIM-SM:985135773293:000}\\ 
		Au, 764.0 \cite{OpenKIM-SM:113843830602:000a,OpenKIM-SM:113843830602:000}\\ 
		Au, 803.0 \cite{OpenKIM-MO:559016907324:000a,OpenKIM-MO:559016907324:000b,OpenKIM-MO:559016907324:000,OpenKIM-MD:120291908751:005}\\ 
		Au, 804.0 \cite{OpenKIM-MO:087738844640:000a,OpenKIM-MO:087738844640:000,OpenKIM-MD:120291908751:005}\\ 
		Au, 811.0 \cite{OpenKIM-MO:557267801129:000a,OpenKIM-MO:557267801129:000,OpenKIM-MD:120291908751:005}\\ 
		Au, 814.0 \cite{OpenKIM-MO:426403318662:000a,OpenKIM-MO:426403318662:000,OpenKIM-MD:120291908751:005}\\ 
		Au, 829.0 \cite{OpenKIM-MO:684444719999:000a,OpenKIM-MO:684444719999:000b,OpenKIM-MO:684444719999:000,OpenKIM-MD:120291908751:005}\\ 
		Au, 832.0 \cite{OpenKIM-MO:318213562153:000a,OpenKIM-MO:318213562153:000,OpenKIM-MD:120291908751:005}\\ 
		Au, 833.0 \cite{OpenKIM-MO:592431957881:000a,OpenKIM-MO:592431957881:000b,OpenKIM-MO:592431957881:000,OpenKIM-MD:120291908751:005}\\ 
		Au, 836.0 \cite{OpenKIM-MO:468407568810:005a,OpenKIM-MO:468407568810:005b,OpenKIM-MO:468407568810:005,OpenKIM-MD:120291908751:005}\\ 
		Au, 847.0 \cite{OpenKIM-SM:066295357485:000a,OpenKIM-SM:066295357485:000}\\ 
		Au, 891.0 \cite{OpenKIM-MO:228280943430:000a,OpenKIM-MO:228280943430:000,OpenKIM-MD:120291908751:005}\\ 
		Au, 991.0 \cite{OpenKIM-MO:188701096956:000a,OpenKIM-MO:188701096956:000,OpenKIM-MD:120291908751:005}\\ 
		Au, 1253.0 \cite{OpenKIM-MO:173248269481:000a,OpenKIM-MO:173248269481:000,OpenKIM-MD:120291908751:005}\\ 
		Au, 1300.0 \cite{OpenKIM-MO:946831081299:000a,OpenKIM-MO:946831081299:000,OpenKIM-MD:120291908751:005}\\ 
		Au, 1345.0 \cite{OpenKIM-MO:754413982908:000a,OpenKIM-MO:754413982908:000,OpenKIM-MD:120291908751:005}\\ 
		Au, 1375.0 \cite{OpenKIM-MO:104891429740:005a,OpenKIM-MO:104891429740:005b,OpenKIM-MO:104891429740:005c,OpenKIM-MO:104891429740:005,OpenKIM-MD:120291908751:005}\\ 
		Cu, 773.0 \cite{OpenKIM-MO:270337113239:005a,OpenKIM-MO:270337113239:005,OpenKIM-MD:120291908751:005}\\ 
		Cu, 1047.0 \cite{OpenKIM-MO:529419924683:000a,OpenKIM-MO:529419924683:000,OpenKIM-MD:536750310735:000}\\ 
		Cu, 1064.0 \cite{OpenKIM-MO:127245782811:005a,OpenKIM-MO:127245782811:005b,OpenKIM-MO:127245782811:005,OpenKIM-MD:120291908751:005}\\ 
		Cu, 1072.0 \cite{OpenKIM-SM:667696763561:000a,OpenKIM-SM:667696763561:000}\\ 
		Cu, 1082.0 \cite{OpenKIM-MO:759493141826:000a,OpenKIM-MO:759493141826:000b,OpenKIM-MO:759493141826:000,OpenKIM-MD:120291908751:005}\\ 
		Cu, 1086.0 \cite{OpenKIM-MO:318213562153:000a,OpenKIM-MO:318213562153:000,OpenKIM-MD:120291908751:005}\\ 
		Cu, 1091.0 \cite{OpenKIM-MO:547744193826:000a,OpenKIM-MO:547744193826:000b,OpenKIM-MO:547744193826:000,OpenKIM-MD:120291908751:005}\\ 
		Cu, 1095.0 \cite{OpenKIM-SM:399364650444:000a,OpenKIM-SM:399364650444:000}\\ 
		Cu, 1097.0 \cite{OpenKIM-MO:950828638160:000a,OpenKIM-MO:950828638160:000b,OpenKIM-MO:950828638160:000,OpenKIM-MD:120291908751:005}\\ 
		Cu, 1097.0 \cite{OpenKIM-MO:380822813353:000a,OpenKIM-MO:380822813353:000b,OpenKIM-MO:380822813353:000,OpenKIM-MD:120291908751:005}\\ 
		Cu, 1143.0 \cite{OpenKIM-MO:426403318662:000a,OpenKIM-MO:426403318662:000,OpenKIM-MD:120291908751:005}\\ 
		Cu, 1143.0 \cite{OpenKIM-MO:469343973171:005a,OpenKIM-MO:469343973171:005,OpenKIM-MD:120291908751:005}\\ 
		Cu, 1145.0 \cite{OpenKIM-MO:346334655118:005a,OpenKIM-MO:346334655118:005,OpenKIM-MD:120291908751:005}\\ 
		Cu, 1149.0 \cite{OpenKIM-MO:128703483589:005a,OpenKIM-MO:128703483589:005,OpenKIM-MD:120291908751:005}\\ 
		Cu, 1151.0 \cite{OpenKIM-MO:666348409573:004a,OpenKIM-MO:666348409573:004b,OpenKIM-MO:666348409573:004,OpenKIM-MD:120291908751:005}\\ 
		Cu, 1157.0 \cite{OpenKIM-MO:266134052596:000a,OpenKIM-MO:266134052596:000,OpenKIM-MD:120291908751:005}\\ 
		Cu, 1159.0 \cite{OpenKIM-MO:020851069572:000a,OpenKIM-MO:020851069572:000,OpenKIM-MD:120291908751:005}\\ 
		Cu, 1161.0 \cite{OpenKIM-MO:119135752160:005a,OpenKIM-MO:119135752160:005,OpenKIM-MD:120291908751:005}\\ 
		Cu, 1168.0 \cite{OpenKIM-MO:592013496703:005a,OpenKIM-MO:592013496703:005,OpenKIM-MD:120291908751:005}\\ 
		Cu, 1172.0 \cite{OpenKIM-MO:145873824897:000a,OpenKIM-MO:145873824897:000,OpenKIM-MD:120291908751:005}\\ 
		Cu, 1189.0 \cite{OpenKIM-MO:931672895580:000a,OpenKIM-MO:931672895580:000,OpenKIM-MD:536750310735:000}\\ 
		Cu, 1190.0 \cite{OpenKIM-MO:097471813275:000a,OpenKIM-MO:097471813275:000,OpenKIM-MD:120291908751:005}\\ 
		Cu, 1200.0 \cite{OpenKIM-MO:831121933939:000a,OpenKIM-MO:831121933939:000,OpenKIM-MD:120291908751:005}\\ 
		Cu, 1209.0 \cite{OpenKIM-SM:316120381362:001a,OpenKIM-SM:316120381362:001}\\ 
		Cu, 1248.0 \cite{OpenKIM-MO:942551040047:005a,OpenKIM-MO:942551040047:005,OpenKIM-MD:120291908751:005}\\ 
		Cu, 1278.0 \cite{OpenKIM-MO:265210066873:000a,OpenKIM-MO:265210066873:000,OpenKIM-MD:536750310735:000}\\ 
		Cu, 1280.0 \cite{OpenKIM-SM:656517352485:000a,OpenKIM-SM:656517352485:000}\\ 
		Cu, 1389.0 \cite{OpenKIM-MO:748636486270:005a,OpenKIM-MO:748636486270:005,OpenKIM-MD:120291908751:005}\\ 
		Cu, 1400.0 \cite{OpenKIM-MO:803527979660:000a,OpenKIM-MO:803527979660:000,OpenKIM-MD:120291908751:005}\\ 
		Cu, 1402.0 \cite{OpenKIM-MO:657255834688:000a,OpenKIM-MO:657255834688:000,OpenKIM-MD:120291908751:005}\\ 
		Cu, 1457.0 \cite{OpenKIM-MO:945691923444:005a,OpenKIM-MO:945691923444:005,OpenKIM-MD:120291908751:005}\\ 
		Cu, 1471.0 \cite{OpenKIM-MO:600021860456:005a,OpenKIM-MO:600021860456:005,OpenKIM-MD:120291908751:005}\\ 
		Cu, 1473.0 \cite{OpenKIM-MO:609260676108:000a,OpenKIM-MO:609260676108:000,OpenKIM-MD:120291908751:005}\\ 
		Cu, 1482.0 \cite{OpenKIM-MO:120596890176:005a,OpenKIM-MO:120596890176:005,OpenKIM-MD:120291908751:005}\\ 
		Cu, 1569.0 \cite{OpenKIM-MO:762798677854:000a,OpenKIM-MO:762798677854:000,OpenKIM-MD:120291908751:005}\\ 
		Cu, 1773.0 \cite{OpenKIM-MO:179025990738:005a,OpenKIM-MO:179025990738:005,OpenKIM-MD:120291908751:005}\\ 
		Cu, 1842.0 \cite{OpenKIM-MO:642748370624:000a,OpenKIM-MO:642748370624:000,OpenKIM-MD:120291908751:005}\\ 
		Cu, 1952.0 \cite{OpenKIM-SM:239791545509:000a,OpenKIM-SM:239791545509:000}\\ 
		Cu, 3554.0 \cite{OpenKIM-SM:521856783904:000}\\ 
		Ni, 1622.0 \cite{OpenKIM-SM:559286646876:000a,OpenKIM-SM:559286646876:000}\\ 
		Ni, 1644.0 \cite{OpenKIM-SM:306597220004:000a,OpenKIM-SM:306597220004:000}\\ 
		Ni, 1748.0 \cite{OpenKIM-MO:776437554506:000a,OpenKIM-MO:776437554506:000,OpenKIM-MD:120291908751:005}\\ 
		Ni, 1761.0 \cite{OpenKIM-MO:592013496703:005a,OpenKIM-MO:592013496703:005,OpenKIM-MD:120291908751:005}\\ 
		Ni, 1790.0 \cite{OpenKIM-MO:110256178378:005a,OpenKIM-MO:110256178378:005b,OpenKIM-MO:110256178378:005,OpenKIM-MD:120291908751:005}\\ 
		Ni, 1813.0 \cite{OpenKIM-MO:047308317761:000a,OpenKIM-MO:047308317761:000,OpenKIM-MD:120291908751:005}\\ 
		Ni, 1826.0 \cite{OpenKIM-MO:593762436933:000a,OpenKIM-MO:593762436933:000b,OpenKIM-MO:593762436933:000,OpenKIM-MD:120291908751:005}\\ 
		Ni, 1858.0 \cite{OpenKIM-MO:222110751402:000a,OpenKIM-MO:222110751402:000,OpenKIM-MD:120291908751:005}\\ 
		Ni, 1875.0 \cite{OpenKIM-MO:832600236922:005a,OpenKIM-MO:832600236922:005,OpenKIM-MD:120291908751:005}\\ 
		Ni, 1881.0 \cite{OpenKIM-MO:306032198193:000a,OpenKIM-MO:306032198193:000,OpenKIM-MD:120291908751:005}\\ 
		Ni, 1894.0 \cite{OpenKIM-SM:477692857359:000a,OpenKIM-SM:477692857359:000}\\ 
		Ni, 1907.0 \cite{OpenKIM-MO:149104665840:005a,OpenKIM-MO:149104665840:005,OpenKIM-MD:120291908751:005}\\ 
		Ni, 1931.0 \cite{OpenKIM-MO:263593395744:000a,OpenKIM-MO:263593395744:000,OpenKIM-MD:536750310735:000}\\ 
		Ni, 1936.0 \cite{OpenKIM-MO:036303866285:000a,OpenKIM-MO:036303866285:000,OpenKIM-MD:120291908751:005}\\ 
		Ni, 1960.0 \cite{OpenKIM-SM:333792531460:001a,OpenKIM-SM:333792531460:001}\\ 
		Ni, 1969.0 \cite{OpenKIM-MO:418978237058:005a,OpenKIM-MO:418978237058:005b,OpenKIM-MO:418978237058:005,OpenKIM-MD:120291908751:005}\\ 
		Ni, 2017.0 \cite{OpenKIM-MO:469343973171:005a,OpenKIM-MO:469343973171:005,OpenKIM-MD:120291908751:005}\\ 
		Ni, 2021.0 \cite{OpenKIM-MO:922363340570:000, OpenKIM-MD:120291908751:005}\\ 
		Ni, 2056.0 \cite{OpenKIM-MO:267721408934:005a,OpenKIM-MO:267721408934:005,OpenKIM-MD:120291908751:005}\\ 
		Ni, 2069.0 \cite{OpenKIM-MO:010059867259:000a,OpenKIM-MO:010059867259:000,OpenKIM-MD:120291908751:005}\\ 
		Ni, 2072.0 \cite{OpenKIM-MO:109933561507:005a,OpenKIM-MO:109933561507:005,OpenKIM-MD:120291908751:005}\\ 
		Ni, 2102.0 \cite{OpenKIM-MO:101214310689:005a,OpenKIM-MO:101214310689:005,OpenKIM-MD:120291908751:005}\\ 
		Ni, 2108.0 \cite{OpenKIM-MO:535504325462:000a,OpenKIM-MO:535504325462:003,OpenKIM-MD:120291908751:005}\\ 
		Ni, 2111.0 \cite{OpenKIM-MO:266134052596:000a,OpenKIM-MO:266134052596:000,OpenKIM-MD:120291908751:005}\\ 
		Ni, 2118.0 \cite{OpenKIM-MO:010613863288:000a,OpenKIM-MO:010613863288:000,OpenKIM-MD:120291908751:005}\\ 
		Ni, 2121.0 \cite{OpenKIM-MO:871937946490:000a,OpenKIM-MO:871937946490:000,OpenKIM-MD:120291908751:005}\\ 
		Ni, 2122.0 \cite{OpenKIM-MO:532072268679:000a,OpenKIM-MO:532072268679:000,OpenKIM-MD:120291908751:005}\\ 
		Ni, 2124.0 \cite{OpenKIM-MO:751354403791:005a,OpenKIM-MO:751354403791:005,OpenKIM-MD:120291908751:005}\\ 
		Ni, 2124.0 \cite{OpenKIM-MO:826591359508:000a,OpenKIM-MO:826591359508:000,OpenKIM-MD:120291908751:005}\\ 
		Ni, 2150.0 \cite{OpenKIM-MO:103383163946:000a,OpenKIM-MO:103383163946:000,OpenKIM-MD:120291908751:005}\\ 
		Ni, 2194.0 \cite{OpenKIM-MO:365106510449:000a,OpenKIM-MO:365106510449:000,OpenKIM-MD:536750310735:000}\\ 
		Ni, 2215.0 \cite{OpenKIM-MO:580571659842:000a,OpenKIM-MO:580571659842:000b,OpenKIM-MO:580571659842:000,OpenKIM-MD:120291908751:005}\\ 
		Ni, 2308.0 \cite{OpenKIM-MO:258836200237:000a,OpenKIM-MO:258836200237:000,OpenKIM-MD:120291908751:005}\\ 
		Ni, 2524.0 \cite{OpenKIM-MO:657255834688:000a,OpenKIM-MO:657255834688:000,OpenKIM-MD:120291908751:005}\\ 
		Ni, 2526.0 \cite{OpenKIM-MO:820335782779:000a,OpenKIM-MO:820335782779:000,OpenKIM-MD:120291908751:005}\\ 
		Ni, 2546.0 \cite{OpenKIM-MO:803527979660:000a,OpenKIM-MO:803527979660:000,OpenKIM-MD:120291908751:005}\\ 
		Ni, 2550.0 \cite{OpenKIM-MO:400591584784:005a,OpenKIM-MO:400591584784:005,OpenKIM-MD:120291908751:005}\\ 
		Ni, 2632.0 \cite{OpenKIM-MO:763197941039:000a,OpenKIM-MO:763197941039:000,OpenKIM-MD:120291908751:005}\\ 
		Ni, 2647.0 \cite{OpenKIM-MO:468686727341:000a,OpenKIM-MO:468686727341:000,OpenKIM-MD:536750310735:000}\\ 
		Ni, 2766.0 \cite{OpenKIM-SM:971529344487:000a,OpenKIM-SM:971529344487:000}\\ 
		Ni, 2896.0 \cite{OpenKIM-SM:770142935022:000a,OpenKIM-SM:770142935022:000}\\ 
		Ni, 3108.0 \cite{OpenKIM-SM:078420412697:001a,OpenKIM-SM:078420412697:001}\\ 
		Ni, 3631.0 \cite{OpenKIM-MO:977363131043:005a,OpenKIM-MO:977363131043:005,OpenKIM-MD:120291908751:005}\\ 
		Ni, 3858.0 \cite{OpenKIM-MO:715003088863:000a,OpenKIM-MO:715003088863:000,OpenKIM-MD:120291908751:005}\\ 
		Ni, 3928.0 \cite{OpenKIM-MO:677715648236:000a,OpenKIM-MO:677715648236:000,OpenKIM-MD:120291908751:005}\\ 
		Ni, 4136.0 \cite{OpenKIM-MO:769632475533:000a,OpenKIM-MO:769632475533:000,OpenKIM-MD:120291908751:005}\\ 
		Ni, 6258.0 \cite{OpenKIM-SM:168413969663:000}\\ 
		Pb, 308.0 \cite{OpenKIM-MO:988703794028:000a,OpenKIM-MO:988703794028:000b,OpenKIM-MO:988703794028:000,OpenKIM-MD:120291908751:005}\\ 
		Pb, 309.0 \cite{OpenKIM-MO:116920074573:005a,OpenKIM-MO:116920074573:005b,OpenKIM-MO:116920074573:005,OpenKIM-MD:120291908751:005}\\ 
		Pb, 375.0 \cite{OpenKIM-MO:699137396381:005a,OpenKIM-MO:699137396381:005b,OpenKIM-MO:699137396381:005c,OpenKIM-MO:699137396381:005,OpenKIM-MD:120291908751:005}\\ 
		Pb, 383.0 \cite{OpenKIM-MO:119135752160:005a,OpenKIM-MO:119135752160:005,OpenKIM-MD:120291908751:005}\\ 
		Pb, 488.0 \cite{OpenKIM-MO:961101070310:000a,OpenKIM-MO:961101070310:000,OpenKIM-MD:120291908751:005}\\ 
		Pd, 926.0 \cite{OpenKIM-MO:865505436319:000a,OpenKIM-MO:865505436319:000,OpenKIM-MD:120291908751:005}\\ 
		Pd, 965.0 \cite{OpenKIM-MO:114797992931:000a,OpenKIM-MO:114797992931:000,OpenKIM-MD:120291908751:005}\\ 
		Pd, 977.0 \cite{OpenKIM-MO:104806802344:005a,OpenKIM-MO:104806802344:005b,OpenKIM-MO:104806802344:005,OpenKIM-MD:120291908751:005}\\ 
		Pd, 987.0 \cite{OpenKIM-MO:108983864770:005a,OpenKIM-MO:108983864770:005b,OpenKIM-MO:108983864770:005,OpenKIM-MD:120291908751:005}\\ 
		Pd, 1067.0 \cite{OpenKIM-MO:786012902615:000a,OpenKIM-MO:786012902615:000b,OpenKIM-MO:786012902615:000,OpenKIM-MD:120291908751:005}\\ 
		Pd, 1196.0 \cite{OpenKIM-MO:169076431435:000a,OpenKIM-MO:169076431435:000,OpenKIM-MD:120291908751:005}\\ 
		Pd, 1270.0 \cite{OpenKIM-MO:532072268679:000a,OpenKIM-MO:532072268679:000,OpenKIM-MD:120291908751:005}\\ 
		Pd, 1351.0 \cite{OpenKIM-MO:993644691224:000a,OpenKIM-MO:993644691224:000b,OpenKIM-MO:993644691224:000,OpenKIM-MD:120291908751:005}\\ 
		Pd, 1372.0 \cite{OpenKIM-SM:559286646876:000a,OpenKIM-SM:559286646876:000}\\ 
		Pd, 7016.0 \cite{OpenKIM-MO:137572817842:000a,OpenKIM-MO:137572817842:000,OpenKIM-MD:120291908751:005}\\ 
		Pt, 1160.0 \cite{OpenKIM-MO:757342646688:000a,OpenKIM-MO:757342646688:000b,OpenKIM-MO:757342646688:000,OpenKIM-MD:120291908751:005}\\ 
		Pt, 1233.0 \cite{OpenKIM-MO:388062184209:000a,OpenKIM-MO:388062184209:000,OpenKIM-MD:120291908751:005}\\ 
		Pt, 1454.0 \cite{OpenKIM-MO:946831081299:000a,OpenKIM-MO:946831081299:000,OpenKIM-MD:120291908751:005}\\ 
		Pt, 1545.0 \cite{OpenKIM-MO:601539325066:000a,OpenKIM-MO:601539325066:000b,OpenKIM-MO:601539325066:000,OpenKIM-MD:120291908751:005}\\ 
		Pt, 1550.0 \cite{OpenKIM-MO:102190350384:005a,OpenKIM-MO:102190350384:005b,OpenKIM-MO:102190350384:005,OpenKIM-MD:120291908751:005}\\ 
		Rh, 2526.0 \cite{OpenKIM-SM:306597220004:000a,OpenKIM-SM:306597220004:000}\\ 
		Rh, 3364.0 \cite{OpenKIM-SM:066295357485:000a,OpenKIM-SM:066295357485:000}\\ 
		
	\end{multicols}
\end{FlushLeft}

\begin{singlespace}
	\printbibliography[heading=subbibintoc,
					   title=Supplementary Information References]
\end{singlespace}

\end{refcontext}

\end{document}